\documentclass[aps,pra,reprint,showpacs,superscriptaddress,longbibliography]{revtex4-1}
\usepackage{xcolor}
\usepackage{amsfonts}
\usepackage{graphicx}
\usepackage{changes}
\usepackage{amsmath,amsthm}
\usepackage{bm,bbm}
\usepackage{newtxtext,newtxmath}
\usepackage{mathrsfs} 
\usepackage{comment}
\usepackage{braket}

\newcommand{\pro}[1]{\ket{#1}\!\!\:\bra{#1}}
\DeclareMathOperator{\tr}{\mathrm{Tr}}
\DeclareMathOperator{\Span}{\mathrm{span}}

\definecolor{michael}{rgb}{0.7,.3,.5}

\newcommand{\ar}[1]{\textcolor{black}{ #1}}
\newcommand{\AR}[1]{\textcolor{black}{ #1}}

\newtheorem{lem}{Lemma}
\newtheorem{theorem}{Theorem}

\usepackage[colorlinks=true,breaklinks=true,allcolors=blue]{hyperref}
\usepackage{url}

\makeatletter
\let\cat@comma@active\@empty
\makeatother
\makeatletter
\AtBeginDocument{\let\LS@rot\@undefined}
\makeatother

\begin{document}

\title{Robust control of quantum systems by quantum systems}
\author{Thomas Konrad}
\email{konradt@ukzn.ac.za}
\affiliation{School of Chemistry and Physics, University of KwaZulu-Natal, Durban, South Africa}
\affiliation{National Institute of Theoretical Physics (NITheP), UKZN Node, South Africa}
\author{Amy Rouillard}
\affiliation{School of Chemistry and Physics, University of KwaZulu-Natal, Durban, South Africa}
\author{Michael Kastner}
\email{kastner@sun.ac.za}
\affiliation{National Institute for Theoretical Physics (NITheP), Stellenbosch 7600, South Africa}
\affiliation{Institute of Theoretical Physics,  Department of Physics, University of Stellenbosch, Stellenbosch 7600, South Africa}
\author{Hermann Uys}
\affiliation{National Laser Centre, Council for Scientific and Industrial Research, Pretoria, South Africa}
\affiliation{Department of Physics, Stellenbosch University, Stellenbosch, South Africa}

\begin{abstract}
Quantum systems can be controlled by other quantum systems in a reversible way, without any information leaking to the outside of the system--controller compound. Such coherent quantum control is deterministic, is less noisy than measurement-based feedback control, and has potential applications in a variety of quantum technologies, including quantum computation, quantum communication and quantum metrology. Here we introduce a coherent feedback protocol, consisting of a sequence of identical interactions with controlling quantum systems, that steers a quantum system from an arbitrary initial state towards a target state. \ar{We determine the broad class of such coherent feedback channels that achieve convergence to the target state, and then stabilise as well as protect it against noise.} Our results imply that also weak system--controller interactions can counter noise if they occur with suitably high frequency.  We provide an example of a control scheme that does not require knowledge \ar{of} the target state encoded in the controllers, which could be the result of a quantum computation. It thus provides a mechanism for autonomous, purely quantum closed-loop control.      
\end{abstract}


\maketitle

\section{Introduction}
Quantum control is at the basis of quantum technologies that store, communicate and process information in quantum systems \cite{DAlessandro,Nielsen2011}. It also enables the test and exploration of the foundations of quantum physics \cite{Weinfurter2017}. Moreover, the sensitivity of suitably controlled quantum systems such as single atoms, ions, and photons, combined with nonclassical properties like entanglement, can be used to improve the spatial resolution of microscopes \cite{OnoOkamotoTakeuchi13,IsraelRosenSilberberg14}, the accuracy of spectroscopy \cite{Leibfried_etal04} or interferometric techniques \cite{GiovannettiLloydMaccone04} (which in turn permit improved detection of electromagnetic fields or gravitational waves), and for other metrological purposes \cite{GiovannettiLloydMaccone2011}.

Closed-loop quantum feedback control refers to a class of control schemes in which an external controller retrieves information about the state of the quantum system to be controlled and, based on this information, actuates a feedback mechanism on the quantum system, for example by applying external driving fields. The functioning of suitably designed closed-loop control schemes does not require precise knowledge of the initial state of the controlled quantum system, and they have the additional advantage of being robust to noise or other external perturbations. The retrieval of information about the state of the quantum system is usually achieved by measuring an observable. Since the outcome of such a measurement is probabilistic, the control scheme becomes a stochastic dynamical process.

Continuous or sequential unsharp measurements of an observable \cite{Diosi1988, Belavkin1989, Wiseman1993, Carmichael93, Korotkov2001, Audretsch2001} have been found to be particularly beneficial for the purpose of information retrieval, as they limit the disturbance due to measurement backaction \cite{FuchsPeres96} and even enable real-time state monitoring of single quantum systems \cite{Doherty.et.al00, Diosi2006, Oxtoby2008, KonradUys2012, KaterMurch19}.   
Combining such measurements with unitary feedback \cite{Wiseman94,Doherty_etal00,Korotkov2001,Audretsch2001,Geremia06,Dotsenko_etal09} has been experimentally implemented to stabilize photonic qubit states in the presence of noise \cite{Gillett_etal10}, to control the number of photons in a cavity  \cite{Sayrin2011} or to control superconducting qubits \cite{Vijay2012}. 

While the controlled system in all these schemes and experiments is quantum mechanical, the feedback mechanism, being based on measurement, is classical at heart. A fully quantum mechanical version of closed-loop feedback control, called coherent quantum feedback control, has been proposed in Ref.~\cite{Lloyd00} and experimentally implemented in Refs.~\cite{Nelson_etal00,HiroseCappellaro16}. Different from measurement-based quantum feedback control as described in the previous paragraph, here the external controller that retrieves information and provides feedback is also a coherently operating quantum system. Since no measurement is involved, coherent feedback control is neither stochastic nor destructive \cite{Lloyd00}. {It} can be easier to implement experimentally, and specific coherent feedback control schemes are known to be advantageous over measurement-based ones for a number of applications, including the cooling of open optical resonators \cite{HamerlyMabuchi12}, suppression of noise in optomechanical systems \cite{Yang_etal15}, and entanglement control in quantum networks \cite{Hein_etal15}.

In this article we contribute to the foundations of coherent feedback control by identifying operations and protocols that guarantee successful control of a quantum system. 

\section{Results} 
The setting we consider consists of a quantum system that we wish to control (``the system''), augmented by an assembly of control quantum systems (q-controllers) that are coupled sequentially, one at a time, and with identical coupling unitaries, to the system. We prove that, for any target state, there is a continuum of system--controller couplings such that an arbitrary (and possibly unknown) initial state is driven towards the desired target state and, in the presence of noise, stabilised near that state. This establishes initial-state-independent robust coherent feedback control. To facilitate applications of the control scheme, we provide, for an arbitrarily chosen target state, a general scheme as well as explicit examples of system--controller coupling unitaries that guarantee convergence and stability. We illustrate resilience to noise for a coherently controlled single qubit. 


\begin{figure}
\centering
\includegraphics[width=\linewidth]{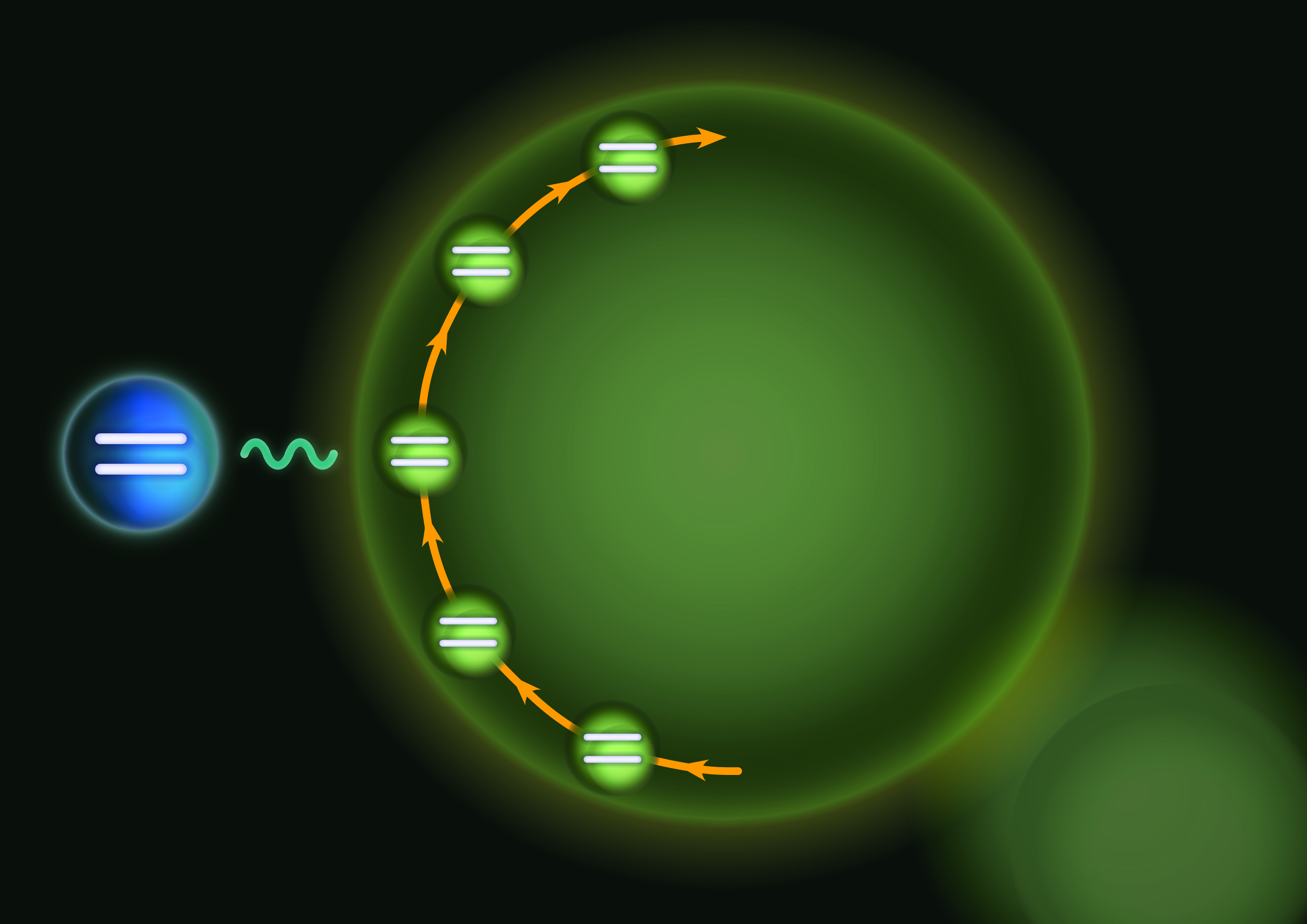}\\[1mm]
\caption{\label{interact}
A (two-level) quantum system (left) is driven gradually by a sequence of interactions with other (two-level) quantum systems (right) into a target state.}
\end{figure}


\subsection{Coherent feedback control with sequential q-controllers} We consider a quantum system $s$ with {states in a } Hilbert space $\mathscr{H}_s$ of finite dimension $d$, and a set of quantum controllers $\{c_1,c_2,\dotsc\}$, each of which has a Hilbert space $\mathscr{H}_{c_m}$ of dimension $n\ge d$. The quantum system is in a (possibly unknown) initial state $\ket{\psi}$
, and each of the q-controllers is prepared in the state $\ket{\psi_0}_{c_m}$. We denote the total initial state, comprising the system and all controllers $c_m$, as
\begin{equation}
\ket{\Psi}\equiv\ket{\psi}\otimes\ket{\psi_0}_{c_1}\otimes\ket{\psi_0}_{c_2}\otimes\dotsb. 
\end{equation}
The quantum system is made to interact sequentially and pairwise with the q-controllers via the unitary  time-evolution operators $U_{s,c_m}$, which generates the mapping
\begin{equation}\label{e:UnitaryMapping}
\ket{\Psi}\mapsto \dotsb U_{s,c_2}U_{s,c_1}\ket{\Psi},
\end{equation}
see Fig.~\ref{interact} for an illustration. Except for the fact that they act on different parts of the total Hilbert space, we assume all $U_{s,c_m}$ to be identical, and thus  omit their indices. The mapping \eqref{e:UnitaryMapping} can be equivalently represented as a repeated (for each q-controller) application of a trace-preserving channel $\$$,
\begin{equation}\label{e:channel}
\rho \mapsto \$(\rho) \equiv\tr_c \left[U({\rho\otimes \rho_c}) U^\dagger\right] = \sum_i M_i \rho M_i^\dagger,
\end{equation} 
where \ar{$\rho_c=\ket{\psi_0}_c{\vphantom{\ket{\psi_0}}}_c\!\bra{\psi_0}$} is the initial state of the q-controller, and the $n$-element set of Kraus operators \cite{Nielsen2011}
\AR{\begin{equation}
M_i = {\vphantom{\ket{i}}}_c\!\braket{i|U|\psi_0}_c,\qquad i=0,\dotsc,{n-1},
\label{KrausU}
\end{equation}}
satisfies 
\begin{equation} \label{complete}
\sum_i M_i^\dagger M_i = \mathbb{I}
\end{equation}
for a basis $(\ket{i}_c)_i$ of $\mathscr{H}_{c}$ . The mapping \eqref{e:channel} acts on the reduced density operator $\rho$ of the system only, while the action of the q-controller has been incorporated into the Kraus operators.

The goal of our control scheme is to reach and stabilize a system target state $\ket{T}\in\mathscr{H}_s$. We assume the target to be a pure state, but mixed target states are possible as well. 
A necessary requirement for the control scheme to reach and stabilize the target state $\ket{T}$ is the fixed point condition
\begin{equation}\label{SFP}
M_i \ket{T}= z_i\ket{T}
\end{equation}    
for all \AR{$i=0,\ldots,n-1$}. Equation (\ref{complete}) implies that the eigenvalues $z_i\in \mathbb{C}$ satisfy 
\begin{equation} \label{z}
\sum_i |z_i|^2 =  \sum_i \braket{T|M_i^\dagger M_i|T} = 1. 
\end{equation}
 
Equation \eqref{SFP} guarantees that, once the target state is reached, the q-controllers do not force the system to leave it. A  more restrictive condition was discovered with different reasoning in the context of a measurement-based quantum feedback control scheme \cite{Uys2018}. 
In addition to the fixed point condition \eqref{SFP}, initial-state-independent convergence to the target state further requires that the fixed point be globally attractive. We will prove that this is the case, provided that the set \AR{$\{M_i^\dagger\ket{T}\}_{i=0,\dotsc,n-1}$} spans $\mathscr{H}_s$.  

\subsection{Proof of convergence}%
As a measure of the overlap between some system state $\rho$ and the target state $\ket{T}$ we introduce the {\em target fidelity}
\begin{equation}\label{e:targetfidelity}
F\left(\rho,T\right) \equiv \braket{T|\rho|T}.
\end{equation}
$F$ can take on values between zero and one, and we have $F=1$ if and only if $\rho=\ket{T}\bra{T}$, { i.e., the system is in} the target state.
\begin{lem} \label{lemma1}
The quantum channel \eqref{e:channel} with Kraus operators $M_i$ satisfying the condition \eqref{SFP} never decreases the target fidelity,
\begin{equation}
\Delta F(\rho) \equiv F\left(\$(\rho),T\right) - F\left(\rho,T\right)\ge 0 . \label{theorem0}
\end{equation}
\end{lem}
\noindent {\it Proof:} 
We prove that $\Delta F$ is nonnegative by expressing it as a sum of squared norms, \begin{equation}\label{proof1}
\Delta F(\rho) =\sum_i \left\lVert A_i \ket{T}\right\rVert^2
\end{equation}
with
\begin{equation}
A_i = \sqrt{\rho} \left(\mathbb{I} - \pro{T}\right)M_i^\dagger.
\end{equation}
The validity of \eqref{proof1} is confirmed by evaluating, with the help of \eqref{SFP}, the summands on its right-hand side, 
\begin{multline}
\braket{T|A _i^\dagger A_i|T} = \braket{T|M_i \rho M_i^\dagger|T} + |z_i|^2 \braket{T|\rho|T}\\
 - \braket{T|\rho M_i^\dagger M_i|T} - \braket{T|M_i^\dagger M_i \rho|T}.
\end{multline}
Summing these terms over $i$ and making use of Eqs.\ \eqref{complete} and \eqref{z} proves the claim \eqref{proof1}.
\qed

Alternatively, the nonnegativity condition \eqref{theorem0} follows from the fixed point condition $\$(\pro{T}) = \pro{T}$ together with the monotonicity of the fidelity under trace-preserving channels \cite{Nielsen2011}.

If the target state is the only fix point of the channel action and the system gains target fidelity for all other states, then the target fidelity will increase in an iterative application of the channel $\$$ until the target state is reached. The following Lemma gives a criterion to check the strict monotonicity of the fidelity.     

\begin{lem} \label{lemma2} The quantum channel \eqref{e:channel} with Kraus operators $M_i$ satisfying the condition \eqref{SFP} strictly increases the target fidelity, $\Delta F(\rho) > 0$, for all states $\rho \neq \pro{T}$ if and only if the vectors { $M_i^\dagger\ket{T}$} span the Hilbert space $\mathscr{H}_s$ of the system. \end{lem}

\noindent {\it Proof:} According to Lemma \ref{lemma1} we have $\Delta F(\rho) \ge 0$. Equality can occur if and only if all vectors on the right-hand side of Eq.~\eqref{proof1} vanish individually, 
\AR{\begin{equation}\label{e:DeltaF0Equivalence}
 \Delta F(\rho) = 0 \;\Longleftrightarrow\; {P M_i^\dagger \ket{T}= 0\quad\forall i=0, \dotsc, n-1,}
\end{equation}}
where {$P=\sqrt{\rho} \left (\mathbb{I} - \pro{T}\right)$.} One way for the right-hand equation in \eqref{e:DeltaF0Equivalence} to hold for all $i$ is if $P$ is the zero operator. However, this possibility requires $\rho = \pro{T}$ and is hence precluded by assumption in Lemma \ref{lemma2}. For $P\neq0$, and if furthermore the vectors {$M_i^\dagger\ket{T}$ span}  $\mathscr{H}_s$, then at least one of them must satisfy {$P M_i^\dagger\ket{T}\neq0$,} which implies $\Delta F(\rho)>0$ by virtue of Eq.~\eqref{proof1}.
If, on the other hand, the vectors $M_i^\dagger\ket{T}$ do not span $\mathscr{H}_s$, then there exists a vector $\ket{\psi_\perp}\neq0$ such that $\bra{\psi_\perp} M_i^\dagger\ket{T} =0$ for all \AR{$i=0,\dotsc,n-1$}. In conjunction with Eqs.\ \eqref{complete} and \eqref{SFP} this implies that $\braket{\psi_\perp|T} = \sum \braket{\psi_\perp|M_i^\dagger M_i|T} = \sum z_i \braket{\psi_\perp|M_i^\dagger|T} =0$. From this property it follows that $\Delta F\left(\ket{\psi_\perp}\bra{\psi_\perp}\right)= 0$. Requiring $\Delta F(\rho)>0$ for all $\rho\neq \pro{T}$ therefore implies that the linear span of \AR{$\{M_i^\dagger\ket{T}\}_{i=0,\dotsc,n-1}$} equals $\mathscr{H}_s$. \qed 

We can now use Lemma~\ref{lemma2} to prove initial-state-independent convergence to the target state under coherent feedback control with sequential and identical q-controllers.   
\begin{theorem}\label{theorem1}
Let $\$$ be a quantum channel as defined in \eqref{e:channel} with Kraus operators $M_i$ {obeying} the condition \eqref{SFP} with respect to some target state $\ket{T}\in \mathscr{H}_s$, and satisfying
\begin{equation}\label{e:span}
{\Span\bigl\{M_i^\dagger  \ket{T}\bigr\}_i=\mathscr{H}_s.}
\end{equation}
Then any state $\rho$ converges to the target state under repeated application of the quantum channel,
\begin{align}\label{claim}
\lim_{n\to\infty} \$^n(\rho)= \pro{T}. 
\end{align}
\end{theorem}
\noindent{\it Proof}: Define $\rho_n:=\$^n(\rho)$. (i) If $\rho_n= \pro{T}$ for some finite $n$, then the fixed point condition $\$^n(\pro{T})= \pro{T}$ implies \eqref{claim}.
(ii) If $\rho_n\neq \pro{T}$ for all finite $n$, then Lemma~\ref{lemma2} implies that $F_n:=F(\rho_n,T)$ is a strictly increasing sequence in $n$. Since $F_n\leq1$ is upper bounded, the sequence converges, and hence $\lim_{n\to\infty} \Delta F (\rho_n)= 0$ with $\Delta F(\rho_n):=F_{n+1} - F_n$. By continuity of $\Delta F$ it follows that
\begin{equation}\label{e:DFlimits}
\Delta F \Bigl(\lim_{n\to\infty}\rho_n\Bigr) = \lim_{n\to\infty} \Delta F (\rho_n)= 0.
\end{equation}
According to Lemma~\ref{lemma2}, $\Delta F(\rho)=0$ implies $\rho=\pro{T}$, and hence Eq.~\eqref{e:DFlimits} implies \eqref{claim}. 
\qed

\subsection{Construction of control channels}
\label{s:construction}
\ar{There are at least two methods} to construct control channels that lead to convergence to a target state $\ket{T}$. Both construction methods use as a starting point a positive-operator valued measure (POVM), i.e., a set of positive operators $E_i$ satisfying $\sum_i E_i = \mathbb{I}$. {The first construction method requires that the POVM additionally satisfies}
\begin{equation}\label{e:spanE}
\Span\bigl\{E_i \ket{T}\bigr\}_i=\mathscr{H}_s.
\end{equation}
An example of POVMs satisfying \eqref{e:spanE} for any target state $\ket{T}$ are informationally complete POVMs, defined as consisting of \ar{elements} $E_i$ that span the space of observables on $\mathscr{H}_s$ \cite{Uys2018}. Another example {is POVMs consisting of pairwise commuting $E_i$ that correspond to} unsharp measurements of nondegenerate observables (see Sec.~A of the Supplementary Information (SI)).  Such POVMs satisfy \eqref{e:spanE} for target states for which \AR{$\ket{T}=\sum_{k=0}^{d-1} t_k\ket{k}$} with $t_k\neq0$ for all $k$, where $\ket{k}$ denote the simultaneous eigenvectors of the commuting $E_i$. To turn such a POVM into a control channel with target state $\ket{T}$, we define Kraus operators 
\begin{equation}
M_i\equiv U_i\sqrt{E_i}
\label{Krausform}
\end{equation} 
in terms of the $E_i$ and choose the unitaries $U_i$ such that the vectors $\sqrt{E_i}\ket{T}$ are rotated onto $\ket{T}$, i.e., $U_i\sqrt{E_i}\ket{T}\propto\ket{T}$ provided that $E_i\ket{T}\neq 0$. It is straightforward to verify that the $M_i$ constructed in this way satisfy \eqref{SFP} and \eqref{e:span}, and hence the conditions of Theorem~\ref{theorem1}.

\ar{The second method to construct a converging control channel also employs commutative POVMs corresponding to unsharp measurements of non-degenerate observables. However, here we consider targets that are a superposition of only $m<d \equiv \operatorname{dim}(\mathcal{H}_s)$ eigenvectors of the POVM, \AR{$\ket{T}=\sum_{k=0}^{m-1} t_k\ket{k}$}. In addition, only  $m$ POVM elements satisfy $E_i\ket{T}\not=0$ and as a result, the POVM  violates condition~\eqref{e:spanE} for the target state.} An arbitrary extension \AR{$\ket{T'}=\ket{T} + \sum_{k=m}^{d-1} t_k\ket{k}$} with $t_k\not=0$ however, \ar{leads to } $\Span\bigl\{E_i \ket{T'}\bigr\}_i=\mathscr{H}_s$ and can be used to design a control channel \ar{for} target state $\ket{T}$. We show in Sec.~A of the SI that Kraus operators $M_i=U_i \sqrt{E_i}$ constructed such that $U_i\sqrt{E_i}\ket{T'}\propto\ket{T}$ for $E_i\ket{T}= 0$, and $U_i\sqrt{E_i}\ket{T}\propto\ket{T}$ otherwise, satisfy conditions \eqref{SFP} and (\ref{e:span}).

All channels can be implemented by unitary couplings to additional systems \cite{Kraus83}, which in our case play the role of controllers. In the following we study examples of channels and their unitary realisation that correspond to the two construction  methods described above.   

\begin{figure}
\centering
\includegraphics[width=0.49\linewidth]{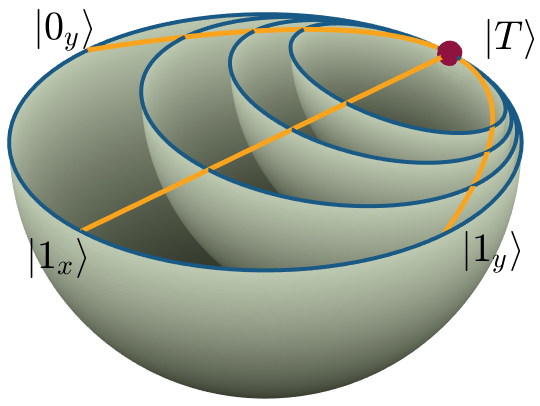}
\includegraphics[width=0.49\linewidth]{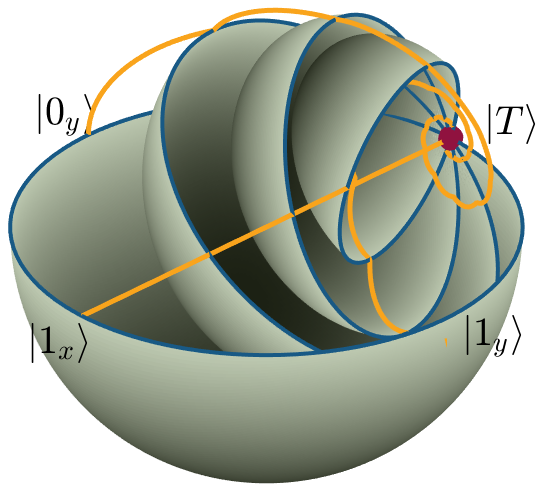}
\caption{\label{trajectories}
Evolution of the Bloch sphere of states, represented by its lower hemisphere, towards the target state \ar{(red point)} under repeated applications of a control channel. The orange lines show the evolution of three states \ar{(eigenstates of Pauli operators)} during the interaction with controllers initialised in the \ar{target state $\ket{0_x}_c$}. Left: For control channel \eqref{e:Kraus} with $\lambda=\pi/5$, realised through system--controller interaction \eqref{U}. Right: For control channel \eqref{channel2} with $\lambda=\pi/5$, realised through interaction \eqref{Uall2}.}
\end{figure}

\subsection{Example 1: \ar{Basic control mechanism}}
\label{s:example1}
{Here we} construct a control channel with a POVM and a target state $\ket{T}$  that satisfy condition (\ref{e:spanE}). For simplicity we consider a {single} qubit, and without loss of generality we choose a basis \AR{$\{\ket{0},\ket{1}\}$} of $\mathscr{H}_s$ such that \AR{$\ket{T}=\left(\ket{0}+\ket{1}\right)/\sqrt{2}$}. One can verify that a one-parameter family of Kraus operators of form \eqref{Krausform} that satisfies conditions \eqref{SFP} and \eqref{e:span} is given by
\AR{\begin{equation}\label{e:Kraus}
\begin{split}
M_0&=e^{-i\sigma_y(\beta-\pi/4)}\left(\sin\beta\pro{0}+\cos\beta\pro{1}\right),\\
M_1&=e^{+i\sigma_y(\beta-\pi/4)}\left(\cos\beta\pro{0}+\sin\beta\pro{1}\right),
\end{split}
\end{equation}}
with \AR{$\sigma_y=i\ket{1}\bra{0}-i\ket{0}\bra{1}$} and $\beta\in\mathbbm{R}$. 
Figure \ref{trajectories} (left) {illustrates} the evolution of states under repeated action of the control channel~\eqref{e:Kraus}. Initial states on the outermost shell, corresponding to the Bloch sphere of pure states, are contracted into the interior of the sphere.
At the same time the contracting spheres shift towards the target state $\ket{T}$, confirming convergence towards the unique fixed point of the dynamics. Orange lines in the plot mark the trajectories of selected initial states during the couplings to the controllers \ar{which} implement the consecutive executions of the channel.

Remarkably, even though initial as well as target states are pure, the control channel takes the system through a sequence of mixed states.
In the special case that the initial state is orthogonal to the target state, the trajectory leads, independently of the choice of parameter $\beta$, on the shortest path through the centre of the Bloch sphere, which represents the maximally mixed state. The control channel specified by (19) can be realised by \ar{consecutive interaction with control systems governed by the unitary dynamics
\begin{equation}\label{U}
U = \exp\left(-\frac{i\lambda}{2} (\sigma_y\otimes \sigma_y + \sigma _z\otimes \sigma_z)\right).
\end{equation}
The controllers are initialised in the target state \AR{$\ket{0_x}_c=\left(\ket{0}_c + i\ket{1}_c\right)/\sqrt{2}$} where the controller basis \AR{$(\ket{0}_c, \ket{1}_c)$} is the eigenbasis of $\sigma_y$. \ar{The basic control mechanism exerted by $U$ becomes obvious for $\lambda=\pi/2$ by decomposing it into a rotation $U_1$ of the controller conditioned on the state of the system,  followed by a rotation $U_2$  of the system conditioned on the state of the controller:
\AR{\begin{align} 
( \alpha\ket{0} + \beta \ket{1})\ket{0_x}_c&\xrightarrow{U_1} \alpha\ket{0}\ket{0}_c + \beta \ket{1}\ket{1}_c \\
&\xrightarrow{U_2} \alpha\ket{0_x}\ket{0}_c + \beta \ket{0_x}\ket{1}_c,
\end{align}} }
leaving the system in the target state \AR{$\ket{0_x}= \left(\ket{0} + \ket{1}\right)/\sqrt{2}$}. Here $U_1\equiv\exp(-\frac{i\pi}{4} (\sigma_z\otimes \sigma_z))$ and $U_2\equiv\exp(-\frac{i\pi}{4} (\sigma_y\otimes \sigma_y))$.}
  
The Kraus operators \eqref{e:Kraus} are {recovered by inserting \eqref{U} into \eqref{KrausU}}. 
In Sec.~B of the SI we show that, {under repeated application of the control channel \eqref{e:Kraus}}, the target fidelity increases \ar{in an exponential fashion} with the number $n$ of interactions\ar{:}
 \begin{equation} \label{targetfidelity2}
F(\$^n(\rho),T)= 1 - (1-F_0)(1-\gamma)^n
\end{equation} 
with $\gamma= \sin^2\lambda$, where $F_0= F(\rho,T)$ is the target fidelity of the initial state $\rho$ of the system.  

{Experimentally, a} unitary evolution {of the form} \eqref{U} {is realised in} quantum dots coupled through a cavity \cite{Imamoglu99}, in superconducting qubits \cite{Siewert00}, or in nuclear spins interacting via an electron gas \cite{Lawrence01}. {In all these cases the interaction is characterised} by a Hamiltonian 
\begin{align}
&H = \frac{\hbar\lambda}{2\delta t}  (\sigma_y\otimes \sigma_y + \sigma _z\otimes \sigma_z),
\label{H}
\end{align} 
where the coupling strength $\lambda$ is scaled by the interaction period $\delta t$.
{A control channel implemented by means of this Hamiltonian} is robust against variations of periods $\delta t$ and coupling strengths $\lambda$ {along} the {sequence} of {system--controller} interactions. Such variations only {affect} the parameter $\gamma$ and thus the increment of the target fidelity $\Delta F = \gamma (1-F)$ during the respective interaction period (cp.\ SI, Sec.\ B). They thus {modify} the convergence speed, but {do} not \ar{impede} the convergence to the target state. \ar{Hamiltonian \eqref{H} can be modified employing a straightforward transformation of the system's basis of the form $H \rightarrow U_s H U_s^\dagger$, in order to reach a different target state $U_s\ket{T}$.}

\subsection{Example 2: \ar{Weak swap}}
In accordance with the second construction method {outlined in Sec.~\ref{s:construction}}, a control channel can be designed with POVMs that {do} not satisfy \eqref{e:spanE}. In contrast to the previous example, all pure (and in fact all mixed) target states of a system with Hilbert space of {finite dimension $d$} can be prepared by iterations of a universal unitary evolution
\begin{align}\label{Uall2}
&U = \exp\left[-i \lambda S \right]
\end{align}  
with swap operator \ar{$S=\sum_{i,j} \ket{i}\bra{j}\otimes  \ket{j}_{c}{\vphantom{\ket{l}}}_c\!\bra{i}$}. Here $(\ket{i})_i$ and  $(\ket{i}_c)_i$ are orthonormal bases of the system's and the q-controller's Hilbert spaces, respectively. The unitary \eqref{Uall2} can be rewritten as $U =  \cos(\lambda) \mathbb{I} - i \sin(\lambda) S$, and it hence creates a superposition of the input states of system and q-controller, $\ket{\psi_0}\otimes \ket{\phi_0}_c$, and a swapped version of it, i.e.,  $S(\ket{\psi_0}\otimes \ket{\phi_0}_c) \equiv \ket{\phi_0}\otimes \ket{\psi_0}_c$. We therefore call $U$ a weak swap, and its operation leaves the system entangled with the q-controller. \ar{Tracing over the degrees of freedom of the q-controller initialised in the target state \AR{$\ket{\psi_0}_c \equiv \ket{0}_c \equiv\ket{T}_c$}, we obtain the Kraus operators, $M_i={}_c\!\bra{i}U\ket{T}_c$ written in the form \eqref{Krausform}:}
\AR{\begin{align}\label{channel2}
\begin{split}
M_0&=e^{-i\lambda \pro{T}}\left[\cos\lambda\, \mathbb{I} +(1- \cos\lambda)\pro{T}\right]\\ 
M_i&= e^{-i\pi(\ket{T}\bra{i} + \ket{i}\bra{T})/2} \sin\lambda \pro{i},\quad i=1,\ldots,d-1. 
\end{split}
\end{align}} 
These Kraus operators satisfy conditions \eqref{SFP} and \eqref{e:span} with respect to the target state \AR{$\ket{T}=\ket{0}$}, and the system therefore converges to $\ket{T}$ upon repeated application of the corresponding quantum channel, even for weak swap operations ($\lambda < \pi/2$). Moreover, the target fidelity increases exponentially with the number of iterations (cp.~SI, Sec.~C), as in the previous example \eqref{targetfidelity2}.

Figure \ref{trajectories} (right) shows, for the qubit case {$d=2$}, the contractions {and rotations} of the Bloch sphere under repeated action of the quantum channel \eqref{channel2} corresponding to the unitary evolution \eqref{Uall2}. In the first phase of the dynamics, system and controller become entangled, which results in an increasingly mixed system state.  
Subsequently, when the system state approaches the pure target state, entanglement between system and controller decreases.  

\begin{figure}
\centering
\includegraphics[width= 0.95\linewidth]{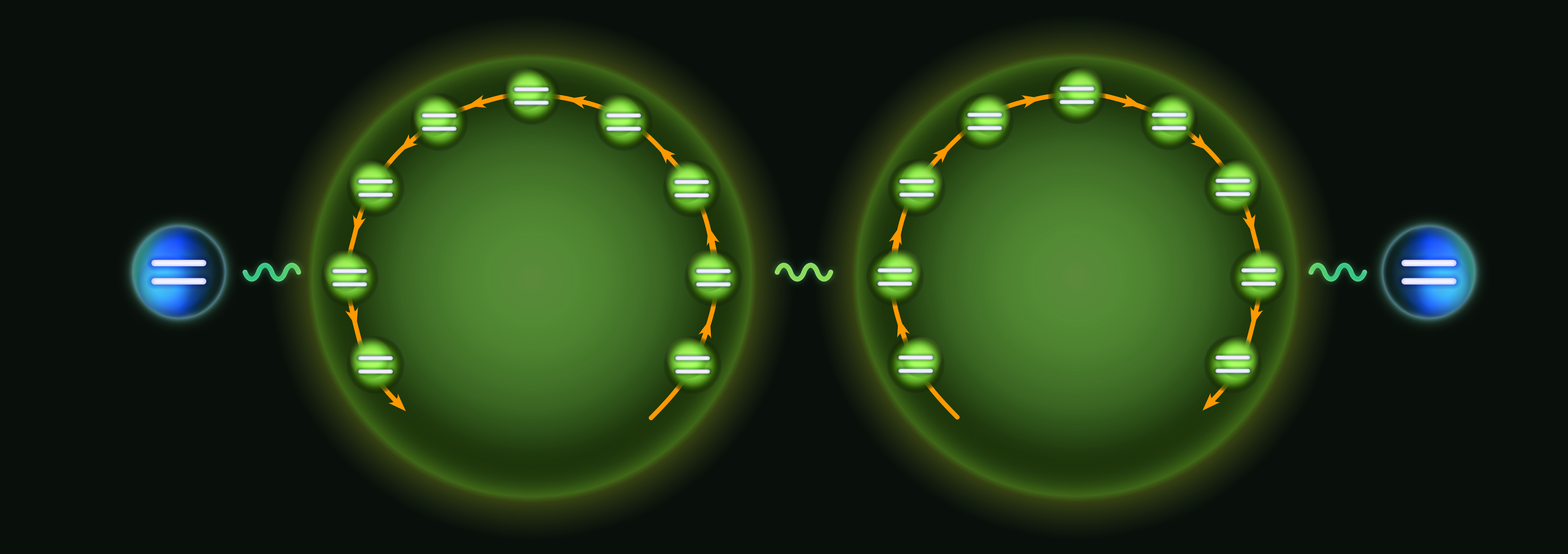}
\caption{\label{qubitpairs} 
Control of a pair of two-level systems (blue) by sequentially coupling each to different controller qubits (green). In spite of {each} controller acting only locally on one of the {sub}systems, target states with nonlocal correlations (entanglement) can be reached and stabilised. For this purpose, the controller qubits need to be entangled first, which can be realised by pairwise interaction \ar{(twiddle in the middle). }
}
\end{figure}

A particularly relevant case to consider is that of $N$ system qubits coupled to $N$ controller qubits. In this case, a swap operation can be decomposed into a tensor product of swaps of the individual system--controller qubits (cp.~SI, Sec.~D), and hence the system qubits can be driven towards a target state by a sequence of simultaneous swaps. Simultaneous swaps may be difficult to implement in a quantum control experiment, but we show in Sec.~E of the SI that this problem can be mitigated: Given a separable target state, the tensor product of swaps can be replaced by a sum, corresponding to {\it independent} weak swap unitaries for each qubit pair.  These can be realised (up to a phase-factor) by a Heisenberg Hamiltonian $H= \frac{\hbar\lambda}{2\delta t}\sum_{i\in\{x,y,z\}}  \sigma _{i}\otimes \sigma_{i}$, acting for time periods of length $\delta t$. An example is the coupling between spins, present for quantum dots \cite{Loss98}, nuclear spins \cite{Kane98}, and in spin-resonance transistors \cite{Vrijen00}. Pairwise Heisenberg interaction, if strong enough ($\lambda \approx \pi/2$), can generate any target state, including entangled states (cp.~SI, Sec.~E). The control of a qubit pair by pairs of controllers is illustrated in Fig.~\ref{qubitpairs}.

\begin{figure}\centering
\includegraphics[width=0.48\linewidth]{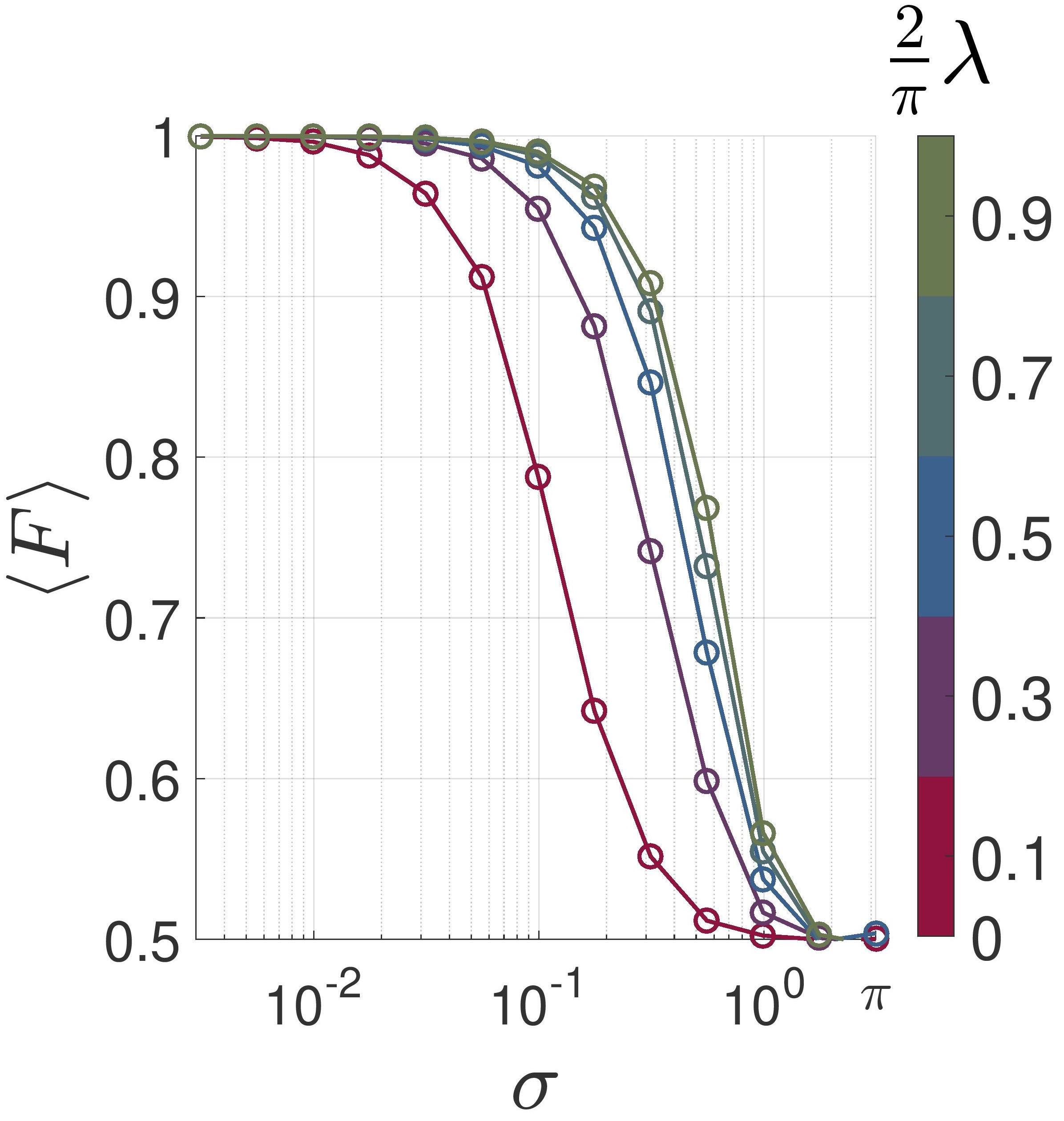}
\hfill
\includegraphics[width=0.48\linewidth]{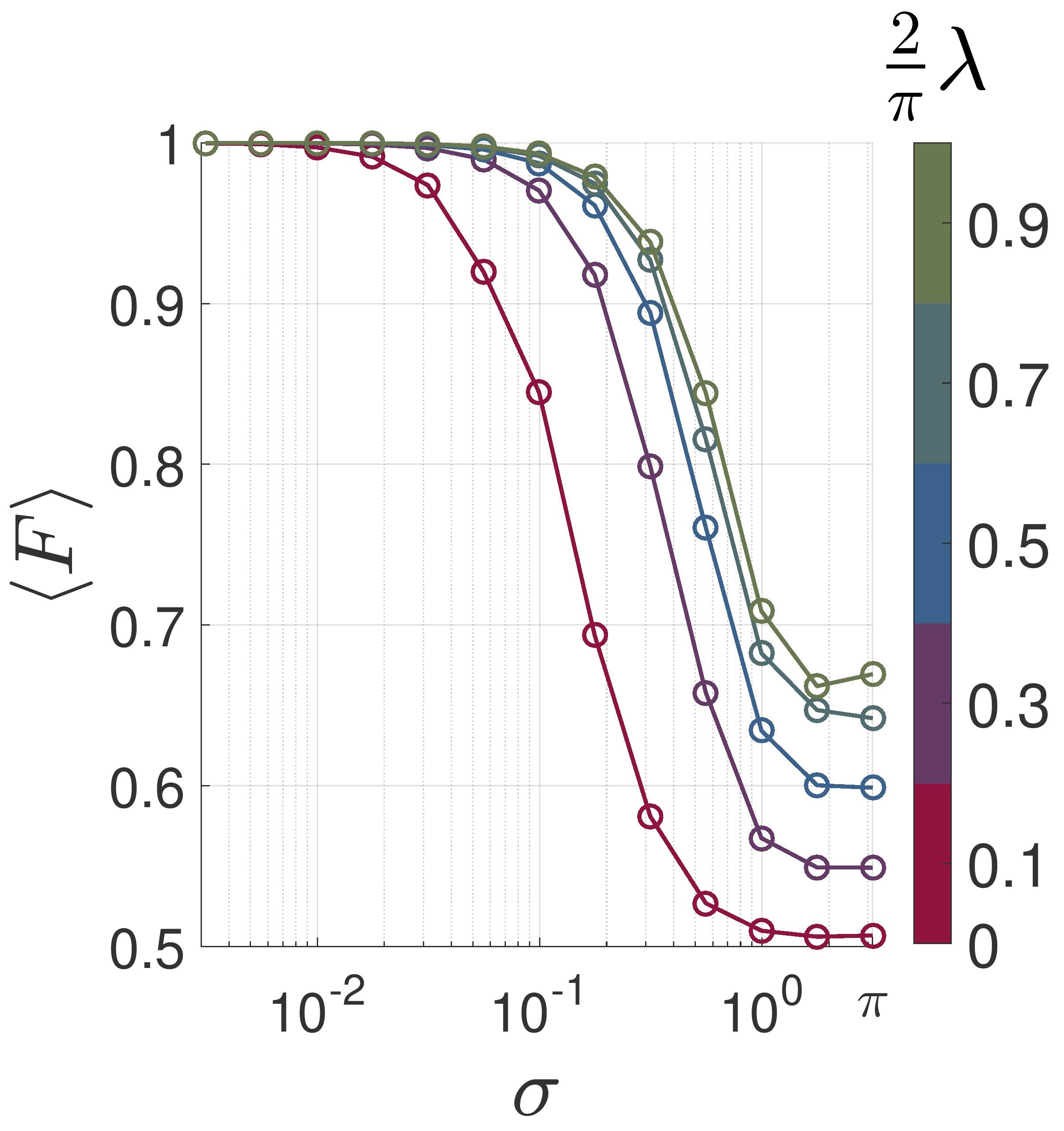}
\caption{Quantitative assessment of the control abilities of the channel \eqref{channel2} in the presence of dephasing noise (left) and depolarisation noise (right). The plots show the averages $\braket{F}:=\sum_{m=1}^N F(\rho_m,T)/N$ of the target fidelity over the sequence $\rho_m$ for trajectories of length $N=10^3$. The value of $\braket{F}$ gives an indication of the size of noise-induced deviations from the target state of the controlled qubit.}
\label{noise}
\end{figure}

To demonstrate and quantify the extent to which repeated applications of the control channel \eqref{channel2} protect a qubit against noise, we consider the simple model $\rho_m=\left(\prod_{k=1}^m R_k\$\right)(\rho)$ in which the control channel $\$$ and realisations $R_k$ of a random unitary are alternatingly applied to the qubit. We consider unitaries $R_k=\exp(-i\theta_k\sigma^{a_k})$ drawn according to the following two distributions. (a) A fixed rotation axis $a_k\equiv z$ for all $k$, with rotation angles $\theta_k$ drawn from a normal distribution with zero mean and standard deviation $\sigma$. This choice models dephasing noise. (b) Rotation axes $a_k\in\{x,y,z\}$ are selected uniformly at random. Rotation angles are chosen normally distributed as in (a). This choice models depolarisation noise.  Figure~\ref{noise} gives a quantitative assessment of the resilience to each of these noise types by plotting the mean {target} fidelity for different noise strengths of the two types of noise. It shows that noise strengths up to $\sigma=\mathscr{O}(10^{-1})$ can successfully be stabilised by control channels of suitable strengths $\lambda$.

\subsection{Example 3: \ar{Entangling qubits}}
\begin{figure}
\includegraphics[width=0.48\linewidth]{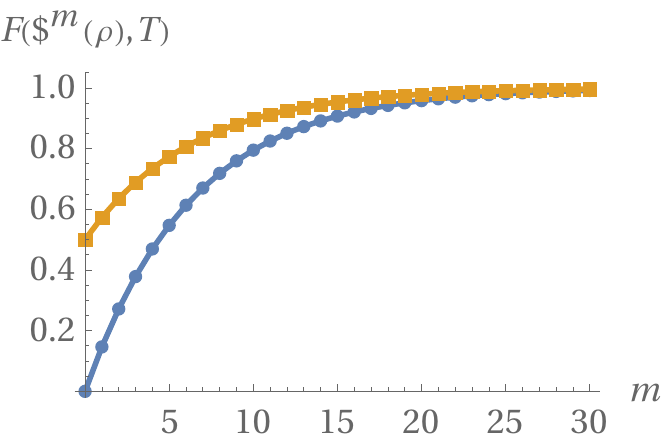}
\includegraphics[width=0.48\linewidth]{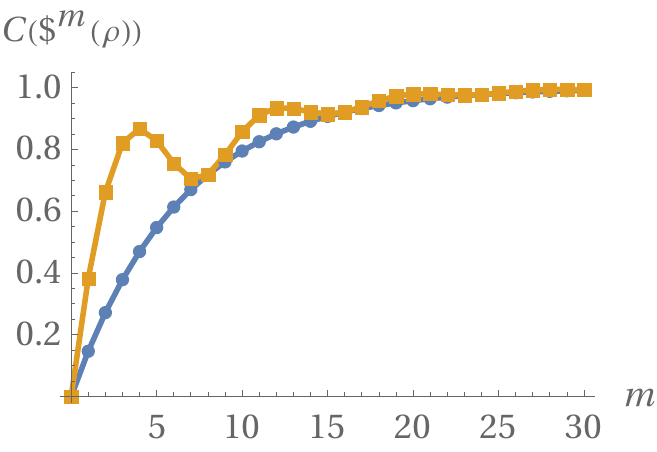}
\caption{Evolution of the target fidelity (left) and the concurrence (right) upon repeated applications of the control channel \ar{of Example 3}. System initial states used are $\ket{\uparrow\uparrow}$ and $\ket{\downarrow\downarrow}$ (orange squares) and $\ket{\uparrow\downarrow}$ and $\ket{\downarrow\uparrow}$ (blue dots).} 
\label{f:Bell}
\end{figure}

In this section we use the control \ar{channel \eqref{channel2}} to entangle two qubits and show that such a channel can be realised using controllers in product states. Unlike other established schemes for quantum state preparation, the control channel has the additional feature of stabilising the \ar{entangled} state against external influences. 
\ar{Applying a suitable transformation $U_s$ on $\mathcal{H}_s$ we obtain a new channel with Kraus operators $\widetilde{M}_i= U_s {M}_i U_s^\dagger$,
where ${M}_i$ is given by \eqref{channel2}. The $\widetilde{M}_i$ are obtained from \eqref{channel2} inserting the Bell basis states}
\begin{subequations}
\AR{\begin{align}
\ket{0}&= (\ket{\uparrow\uparrow}+\ket{\downarrow\downarrow})/\sqrt{2},&\ket{1}&=(\ket{\uparrow\uparrow}-\ket{\downarrow\downarrow})/\sqrt{2},\\
\ket{2}&=(\ket{\uparrow\downarrow}+\ket{\downarrow\uparrow})/\sqrt{2},&\ket{3}&=(\ket{\uparrow\downarrow}-\ket{\downarrow\uparrow})/\sqrt{2}\, .
\end{align}}
\end{subequations}
This control channel enforces convergence towards the Bell target state $\ket{T}\equiv(\ket{\uparrow\uparrow}+\ket{\downarrow\downarrow})/\sqrt{2}$, as illustrated in Fig.~\ref{f:Bell} (left). Using the concurrence \cite{Plenio2014} as a measure, Fig.~\ref{f:Bell} (right) shows that entanglement is built up upon repeated applications of the control channel, in some cases in a nonmonotonic fashion.

\ar{Section~F of the SI contains explicit expressions for the Kraus operators $\widetilde{M}_i$ and the Hamiltonian that realises the control channel with two-qubit controllers  initialised in the state $\ket{\psi_0}_c \equiv \ket{\downarrow\downarrow}_c$.} As a result, entanglement is not just redistributed, but genuinely created in the control process. The system--controller Hamiltonian that achieves this task contains not only pair interactions, but also three- and four-spin operators. While not impossible to deal with in principle, such terms make the experimental implementation of the required unitaries more challenging. In analogy to Sec.~D of the SI, it should be possible to replace the multi-spin interactions by suitably chosen pair interactions, which are more amenable to experimental implementation and would lead, at least approximately, to convergence towards the desired target state.


\section{Discussion}
We proposed a coherent feedback control scheme consisting of repeated applications of identical coherent feedback loops that drive a quantum system into a desired target state $\ket{T}$ and stabilize it in the vicinity of $\ket{T}$. Our main result, Theorem~\ref{theorem1}, establishes controllability in full generality by identifying necessary and sufficient conditions on the control channels such that convergence to $\ket{T}$ is guaranteed for all initial states. This initial-state-independence implies that control channels have the ability to stabilize the target state in the presence of noise, provided the coupling strength and/or the interaction frequency with the controllers are sufficiently high. 

We provide two schemes for the construction of channels satisfying these conditions, and also explicit examples of such channels. While the channels could also be realised by measurement-based closed loop feedback, which is stochastic by nature,  we focus here on deterministic coherent feedback by quantum controllers. We demonstrate coherent control in arbitrary dimensions for the example of consecutive weak swap operations. For single qubits a weak swap can be realised by weak Heisenberg coupling to a quantum controller prepared in the target state. For systems consisting of several qubits, a weak swap may be approximated by Heisenberg couplings between pairs of system qubits and controller qubits, which are more amenable to experimental realisation. Convergence to the desired target state is guaranteed for any strength or interaction period of the coupling unitary, which makes the control channels robust under imperfections in their implementation.

Weak swap operations imprint an arbitrary controller initial state onto the system. Hence, different target states can be reached without modifying the system--controller couplings, but simply by changing the controller initial states. This implies that details of the target state need not be known, as long as an algorithm or a device is available to prepare the controllers in that state. In particular, the target state can be the result of a quantum computation. In this way, the control could be run by quantum systems autonomously.  Alternatively, instead of encoding the desired target state into the controller initial state, a fixed controller initial states may be used, at the expense of having to use a target-specific system--controller coupling as in Examples 1 and 3.

Potential future applications of our coherent quantum control scheme include quantum state preparation, quantum state transfer, and entanglement swapping. A promising direction for future research is a generalization of coherent control to time-evolving targets. Since the proposed control scheme works for arbitrarily weak system--controller couplings, the continuum limit of infinitely frequent, infinitely weak control loops is accessible. Moreover, with the aim of making experimental realizations of control channels more feasible, the construction of optimized approximate system--controller unitaries based only on pair interactions is a desirable objective for more complex quantum information tasks.

\bibliography{xxbib}

\begin{thebibliography}{42}%
\makeatletter
\providecommand \@ifxundefined [1]{%
 \@ifx{#1\undefined}
}%
\providecommand \@ifnum [1]{%
 \ifnum #1\expandafter \@firstoftwo
 \else \expandafter \@secondoftwo
 \fi
}%
\providecommand \@ifx [1]{%
 \ifx #1\expandafter \@firstoftwo
 \else \expandafter \@secondoftwo
 \fi
}%
\providecommand \natexlab [1]{#1}%
\providecommand \enquote  [1]{``#1''}%
\providecommand \bibnamefont  [1]{#1}%
\providecommand \bibfnamefont [1]{#1}%
\providecommand \citenamefont [1]{#1}%
\providecommand \href@noop [0]{\@secondoftwo}%
\providecommand \href [0]{\begingroup \@sanitize@url \@href}%
\providecommand \@href[1]{\@@startlink{#1}\@@href}%
\providecommand \@@href[1]{\endgroup#1\@@endlink}%
\providecommand \@sanitize@url [0]{\catcode `\\12\catcode `\$12\catcode
  `\&12\catcode `\#12\catcode `\^12\catcode `\_12\catcode `\%12\relax}%
\providecommand \@@startlink[1]{}%
\providecommand \@@endlink[0]{}%
\providecommand \url  [0]{\begingroup\@sanitize@url \@url }%
\providecommand \@url [1]{\endgroup\@href {#1}{\urlprefix }}%
\providecommand \urlprefix  [0]{URL }%
\providecommand \Eprint [0]{\href }%
\providecommand \doibase [0]{http://dx.doi.org/}%
\providecommand \selectlanguage [0]{\@gobble}%
\providecommand \bibinfo  [0]{\@secondoftwo}%
\providecommand \bibfield  [0]{\@secondoftwo}%
\providecommand \translation [1]{[#1]}%
\providecommand \BibitemOpen [0]{}%
\providecommand \bibitemStop [0]{}%
\providecommand \bibitemNoStop [0]{.\EOS\space}%
\providecommand \EOS [0]{\spacefactor3000\relax}%
\providecommand \BibitemShut  [1]{\csname bibitem#1\endcsname}%
\let\auto@bib@innerbib\@empty
\bibitem [{\citenamefont {D'Alessandro}(2008)}]{DAlessandro}%
  \BibitemOpen
  \bibfield  {author} {\bibinfo {author} {\bibfnamefont {D.}~\bibnamefont
  {D'Alessandro}},\ }\href {\doibase 10.1201/9781584888833} {\emph {\bibinfo
  {title} {Introduction to Quantum Control and Dynamics}}},\ Applied
  Mathematics and Nonlinear Science Series\ (\bibinfo  {publisher} {Chapman \&
  Hall/CRC, Boca Raton},\ \bibinfo {year} {2008})\BibitemShut {NoStop}%
\bibitem [{\citenamefont {Nielsen}\ and\ \citenamefont
  {Chuang}(2010)}]{Nielsen2011}%
  \BibitemOpen
  \bibfield  {author} {\bibinfo {author} {\bibfnamefont {M.~A.}\ \bibnamefont
  {Nielsen}}\ and\ \bibinfo {author} {\bibfnamefont {I.~L.}\ \bibnamefont
  {Chuang}},\ }\href {\doibase 10.1017/CBO9780511976667} {\emph {\bibinfo
  {title} {Quantum Computation and Quantum Information}}}\ (\bibinfo
  {publisher} {Cambridge University Press, Cambridge},\ \bibinfo {year}
  {2010})\BibitemShut {NoStop}%
\bibitem [{\citenamefont {Rosenfeld}\ \emph {et~al.}(2017)\citenamefont
  {Rosenfeld}, \citenamefont {Burchardt}, \citenamefont {Garthoff},
  \citenamefont {Redeker}, \citenamefont {Ortegel}, \citenamefont {Rau},\ and\
  \citenamefont {Weinfurter}}]{Weinfurter2017}%
  \BibitemOpen
  \bibfield  {author} {\bibinfo {author} {\bibfnamefont {W.}~\bibnamefont
  {Rosenfeld}}, \bibinfo {author} {\bibfnamefont {D.}~\bibnamefont
  {Burchardt}}, \bibinfo {author} {\bibfnamefont {R.}~\bibnamefont {Garthoff}},
  \bibinfo {author} {\bibfnamefont {K.}~\bibnamefont {Redeker}}, \bibinfo
  {author} {\bibfnamefont {N.}~\bibnamefont {Ortegel}}, \bibinfo {author}
  {\bibfnamefont {M.}~\bibnamefont {Rau}}, \ and\ \bibinfo {author}
  {\bibfnamefont {H.}~\bibnamefont {Weinfurter}},\ }\bibfield  {title}
  {\enquote {\bibinfo {title} {Event-ready {B}ell test using entangled atoms
  simultaneously closing detection and locality loopholes},}\ }\href {\doibase
  10.1103/PhysRevLett.119.010402} {\bibfield  {journal} {\bibinfo  {journal}
  {Phys. Rev. Lett.}\ }\textbf {\bibinfo {volume} {119}},\ \bibinfo {pages}
  {010402} (\bibinfo {year} {2017})}\BibitemShut {NoStop}%
\bibitem [{\citenamefont {Ono}\ \emph {et~al.}(2013)\citenamefont {Ono},
  \citenamefont {Okamoto},\ and\ \citenamefont
  {Takeuchi}}]{OnoOkamotoTakeuchi13}%
  \BibitemOpen
  \bibfield  {author} {\bibinfo {author} {\bibfnamefont {T.}~\bibnamefont
  {Ono}}, \bibinfo {author} {\bibfnamefont {R.}~\bibnamefont {Okamoto}}, \ and\
  \bibinfo {author} {\bibfnamefont {S.}~\bibnamefont {Takeuchi}},\ }\bibfield
  {title} {\enquote {\bibinfo {title} {An entanglement-enhanced microscope},}\
  }\href {\doibase 10.1038/ncomms3426} {\bibfield  {journal} {\bibinfo
  {journal} {Nat. Commun.}\ }\textbf {\bibinfo {volume} {4}},\ \bibinfo {pages}
  {2426} (\bibinfo {year} {2013})}\BibitemShut {NoStop}%
\bibitem [{\citenamefont {Israel}\ \emph {et~al.}(2014)\citenamefont {Israel},
  \citenamefont {Rosen},\ and\ \citenamefont
  {Silberberg}}]{IsraelRosenSilberberg14}%
  \BibitemOpen
  \bibfield  {author} {\bibinfo {author} {\bibfnamefont {Y.}~\bibnamefont
  {Israel}}, \bibinfo {author} {\bibfnamefont {S.}~\bibnamefont {Rosen}}, \
  and\ \bibinfo {author} {\bibfnamefont {Y.}~\bibnamefont {Silberberg}},\
  }\bibfield  {title} {\enquote {\bibinfo {title} {Supersensitive polarization
  microscopy using {NOON} states of light},}\ }\href {\doibase
  10.1103/PhysRevLett.112.103604} {\bibfield  {journal} {\bibinfo  {journal}
  {Phys. Rev. Lett.}\ }\textbf {\bibinfo {volume} {112}},\ \bibinfo {pages}
  {103604} (\bibinfo {year} {2014})}\BibitemShut {NoStop}%
\bibitem [{\citenamefont {Leibfried}\ \emph {et~al.}(2004)\citenamefont
  {Leibfried}, \citenamefont {Barrett}, \citenamefont {Schaetz}, \citenamefont
  {Britton}, \citenamefont {Chiaverini}, \citenamefont {Itano}, \citenamefont
  {Jost}, \citenamefont {Langer},\ and\ \citenamefont
  {Wineland}}]{Leibfried_etal04}%
  \BibitemOpen
  \bibfield  {author} {\bibinfo {author} {\bibfnamefont {D.}~\bibnamefont
  {Leibfried}}, \bibinfo {author} {\bibfnamefont {M.~D.}\ \bibnamefont
  {Barrett}}, \bibinfo {author} {\bibfnamefont {T.}~\bibnamefont {Schaetz}},
  \bibinfo {author} {\bibfnamefont {J.}~\bibnamefont {Britton}}, \bibinfo
  {author} {\bibfnamefont {J.}~\bibnamefont {Chiaverini}}, \bibinfo {author}
  {\bibfnamefont {W.~M.}\ \bibnamefont {Itano}}, \bibinfo {author}
  {\bibfnamefont {J.~D.}\ \bibnamefont {Jost}}, \bibinfo {author}
  {\bibfnamefont {C.}~\bibnamefont {Langer}}, \ and\ \bibinfo {author}
  {\bibfnamefont {D.~J.}\ \bibnamefont {Wineland}},\ }\bibfield  {title}
  {\enquote {\bibinfo {title} {Toward {H}eisenberg-limited spectroscopy with
  multiparticle entangled states},}\ }\href {\doibase 10.1126/science.1097576}
  {\bibfield  {journal} {\bibinfo  {journal} {Science}\ }\textbf {\bibinfo
  {volume} {304}},\ \bibinfo {pages} {1476--1478} (\bibinfo {year}
  {2004})}\BibitemShut {NoStop}%
\bibitem [{\citenamefont {Giovannetti}\ \emph {et~al.}(2004)\citenamefont
  {Giovannetti}, \citenamefont {Lloyd},\ and\ \citenamefont
  {Maccone}}]{GiovannettiLloydMaccone04}%
  \BibitemOpen
  \bibfield  {author} {\bibinfo {author} {\bibfnamefont {V.}~\bibnamefont
  {Giovannetti}}, \bibinfo {author} {\bibfnamefont {S.}~\bibnamefont {Lloyd}},
  \ and\ \bibinfo {author} {\bibfnamefont {L.}~\bibnamefont {Maccone}},\
  }\bibfield  {title} {\enquote {\bibinfo {title} {Quantum-enhanced
  measurements: Beating the standard quantum limit},}\ }\href {\doibase
  10.1126/science.1104149} {\bibfield  {journal} {\bibinfo  {journal}
  {Science}\ }\textbf {\bibinfo {volume} {306}},\ \bibinfo {pages} {1330--1336}
  (\bibinfo {year} {2004})}\BibitemShut {NoStop}%
\bibitem [{\citenamefont {Giovannetti}\ \emph {et~al.}(2011)\citenamefont
  {Giovannetti}, \citenamefont {Lloyd},\ and\ \citenamefont
  {Maccone}}]{GiovannettiLloydMaccone2011}%
  \BibitemOpen
  \bibfield  {author} {\bibinfo {author} {\bibfnamefont {V.}~\bibnamefont
  {Giovannetti}}, \bibinfo {author} {\bibfnamefont {S.}~\bibnamefont {Lloyd}},
  \ and\ \bibinfo {author} {\bibfnamefont {L.}~\bibnamefont {Maccone}},\
  }\bibfield  {title} {\enquote {\bibinfo {title} {Advances in quantum
  metrology},}\ }\href {\doibase 10.1038/nphoton.2011.35} {\bibfield  {journal}
  {\bibinfo  {journal} {Nat. Photonics}\ }\textbf {\bibinfo {volume} {5}},\
  \bibinfo {pages} {222--229} (\bibinfo {year} {2011})}\BibitemShut {NoStop}%
\bibitem [{\citenamefont {Di\'{o}si}(1988)}]{Diosi1988}%
  \BibitemOpen
  \bibfield  {author} {\bibinfo {author} {\bibfnamefont {L.}~\bibnamefont
  {Di\'{o}si}},\ }\bibfield  {title} {\enquote {\bibinfo {title} {Continuous
  quantum measurement and {I}t\^{o} formalism},}\ }\href {\doibase
  10.1016/0375-9601(88)90309-X} {\bibfield  {journal} {\bibinfo  {journal}
  {Phys. Lett. A}\ }\textbf {\bibinfo {volume} {129}},\ \bibinfo {pages}
  {419--423} (\bibinfo {year} {1988})}\BibitemShut {NoStop}%
\bibitem [{\citenamefont {Belavkin}(1989)}]{Belavkin1989}%
  \BibitemOpen
  \bibfield  {author} {\bibinfo {author} {\bibfnamefont {V.~P.}\ \bibnamefont
  {Belavkin}},\ }\bibfield  {title} {\enquote {\bibinfo {title} {Nondemolition
  measurement, nonlinear filtering and dynamic programming of quantum
  stochastic processes},}\ }in\ \href {\doibase 10.1007/BFb0041197} {\emph
  {\bibinfo {booktitle} {Modeling and Control of Systems}}},\ \bibinfo {series}
  {Lecture Notes in Control and Information Sciences}, Vol.\ \bibinfo {volume}
  {121},\ \bibinfo {editor} {edited by\ \bibinfo {editor} {\bibfnamefont
  {A.}~\bibnamefont {Blaqui\'{e}re}}}\ (\bibinfo  {publisher} {Springer,
  Berlin},\ \bibinfo {year} {1989})\ pp.\ \bibinfo {pages}
  {245--265}\BibitemShut {NoStop}%
\bibitem [{\citenamefont {Wiseman}\ and\ \citenamefont
  {Milburn}(1993)}]{Wiseman1993}%
  \BibitemOpen
  \bibfield  {author} {\bibinfo {author} {\bibfnamefont {H.~M.}\ \bibnamefont
  {Wiseman}}\ and\ \bibinfo {author} {\bibfnamefont {G.~J.}\ \bibnamefont
  {Milburn}},\ }\bibfield  {title} {\enquote {\bibinfo {title} {Quantum theory
  of field-quadrature measurements},}\ }\href {\doibase
  10.1103/PhysRevA.47.642} {\bibfield  {journal} {\bibinfo  {journal} {Phys.
  Rev. A}\ }\textbf {\bibinfo {volume} {47}},\ \bibinfo {pages} {642--662}
  (\bibinfo {year} {1993})}\BibitemShut {NoStop}%
\bibitem [{\citenamefont {Carmichael}(1993)}]{Carmichael93}%
  \BibitemOpen
  \bibfield  {author} {\bibinfo {author} {\bibfnamefont {H.}~\bibnamefont
  {Carmichael}},\ }\href {\doibase 10.1007/978-3-540-47620-7} {\emph {\bibinfo
  {title} {An Open Systems Approach to Quantum Optics}}}\ (\bibinfo
  {publisher} {Springer, Berlin},\ \bibinfo {year} {1993})\BibitemShut
  {NoStop}%
\bibitem [{\citenamefont {Korotkov}(2001)}]{Korotkov2001}%
  \BibitemOpen
  \bibfield  {author} {\bibinfo {author} {\bibfnamefont {A.~N.}\ \bibnamefont
  {Korotkov}},\ }\bibfield  {title} {\enquote {\bibinfo {title} {Selective
  quantum evolution of a qubit state due to continuous quantum measurement},}\
  }\href {\doibase 10.1103/PhysRevB.63.115403} {\bibfield  {journal} {\bibinfo
  {journal} {Phys. Rev. B}\ }\textbf {\bibinfo {volume} {63}},\ \bibinfo
  {pages} {115403} (\bibinfo {year} {2001})}\BibitemShut {NoStop}%
\bibitem [{\citenamefont {Audretsch}\ \emph {et~al.}(2001)\citenamefont
  {Audretsch}, \citenamefont {Konrad},\ and\ \citenamefont
  {Scherer}}]{Audretsch2001}%
  \BibitemOpen
  \bibfield  {author} {\bibinfo {author} {\bibfnamefont {J.}~\bibnamefont
  {Audretsch}}, \bibinfo {author} {\bibfnamefont {T.}~\bibnamefont {Konrad}}, \
  and\ \bibinfo {author} {\bibfnamefont {A.}~\bibnamefont {Scherer}},\
  }\bibfield  {title} {\enquote {\bibinfo {title} {Sequence of unsharp
  measurements enabling a real-time visualization of a quantum oscillation},}\
  }\href {\doibase 10.1103/PhysRevA.63.052102} {\bibfield  {journal} {\bibinfo
  {journal} {Phys. Rev. A}\ }\textbf {\bibinfo {volume} {63}},\ \bibinfo
  {pages} {052102} (\bibinfo {year} {2001})}\BibitemShut {NoStop}%
\bibitem [{\citenamefont {Fuchs}\ and\ \citenamefont
  {Peres}(1996)}]{FuchsPeres96}%
  \BibitemOpen
  \bibfield  {author} {\bibinfo {author} {\bibfnamefont {C.~A.}\ \bibnamefont
  {Fuchs}}\ and\ \bibinfo {author} {\bibfnamefont {A.}~\bibnamefont {Peres}},\
  }\bibfield  {title} {\enquote {\bibinfo {title} {Quantum-state disturbance
  versus information gain: Uncertainty relations for quantum information},}\
  }\href {\doibase 10.1103/PhysRevA.53.2038} {\bibfield  {journal} {\bibinfo
  {journal} {Phys. Rev. A}\ }\textbf {\bibinfo {volume} {53}},\ \bibinfo
  {pages} {2038--2045} (\bibinfo {year} {1996})}\BibitemShut {NoStop}%
\bibitem [{\citenamefont {Doherty}\ \emph {et~al.}(1999)\citenamefont
  {Doherty}, \citenamefont {Tan}, \citenamefont {Parkins},\ and\ \citenamefont
  {Walls}}]{Doherty.et.al00}%
  \BibitemOpen
  \bibfield  {author} {\bibinfo {author} {\bibfnamefont {A.~C.}\ \bibnamefont
  {Doherty}}, \bibinfo {author} {\bibfnamefont {S.~M.}\ \bibnamefont {Tan}},
  \bibinfo {author} {\bibfnamefont {A.~S.}\ \bibnamefont {Parkins}}, \ and\
  \bibinfo {author} {\bibfnamefont {D.~F.}\ \bibnamefont {Walls}},\ }\bibfield
  {title} {\enquote {\bibinfo {title} {State determination in continuous
  measurement},}\ }\href {\doibase 10.1103/PhysRevA.60.2380} {\bibfield
  {journal} {\bibinfo  {journal} {Phys. Rev. A}\ }\textbf {\bibinfo {volume}
  {60}},\ \bibinfo {pages} {2380--2392} (\bibinfo {year} {1999})}\BibitemShut
  {NoStop}%
\bibitem [{\citenamefont {Di\'{o}si}\ \emph {et~al.}(2006)\citenamefont
  {Di\'{o}si}, \citenamefont {Konrad}, \citenamefont {Scherer},\ and\
  \citenamefont {Audretsch}}]{Diosi2006}%
  \BibitemOpen
  \bibfield  {author} {\bibinfo {author} {\bibfnamefont {L.}~\bibnamefont
  {Di\'{o}si}}, \bibinfo {author} {\bibfnamefont {T.}~\bibnamefont {Konrad}},
  \bibinfo {author} {\bibfnamefont {A.}~\bibnamefont {Scherer}}, \ and\
  \bibinfo {author} {\bibfnamefont {J.}~\bibnamefont {Audretsch}},\ }\bibfield
  {title} {\enquote {\bibinfo {title} {Coupled {I}t\^{o} equations of
  continuous quantum state measurement and estimation},}\ }\href {\doibase
  10.1088/0305-4470/39/40/L01} {\bibfield  {journal} {\bibinfo  {journal} {J.
  Phys. A}\ }\textbf {\bibinfo {volume} {39}},\ \bibinfo {pages} {L575--L581}
  (\bibinfo {year} {2006})}\BibitemShut {NoStop}%
\bibitem [{\citenamefont {Oxtoby}\ \emph {et~al.}(2008)\citenamefont {Oxtoby},
  \citenamefont {Gambetta},\ and\ \citenamefont {Wiseman}}]{Oxtoby2008}%
  \BibitemOpen
  \bibfield  {author} {\bibinfo {author} {\bibfnamefont {N.~P.}\ \bibnamefont
  {Oxtoby}}, \bibinfo {author} {\bibfnamefont {J.}~\bibnamefont {Gambetta}}, \
  and\ \bibinfo {author} {\bibfnamefont {H.~M.}\ \bibnamefont {Wiseman}},\
  }\bibfield  {title} {\enquote {\bibinfo {title} {Model for monitoring of a
  charge qubit using a radio-frequency quantum point contact including
  experimental imperfections},}\ }\href {\doibase 10.1103/PhysRevB.77.125304}
  {\bibfield  {journal} {\bibinfo  {journal} {Phys. Rev. B}\ }\textbf {\bibinfo
  {volume} {77}},\ \bibinfo {pages} {125304} (\bibinfo {year}
  {2008})}\BibitemShut {NoStop}%
\bibitem [{\citenamefont {Konrad}\ and\ \citenamefont
  {Uys}(2012)}]{KonradUys2012}%
  \BibitemOpen
  \bibfield  {author} {\bibinfo {author} {\bibfnamefont {T.}~\bibnamefont
  {Konrad}}\ and\ \bibinfo {author} {\bibfnamefont {H.}~\bibnamefont {Uys}},\
  }\bibfield  {title} {\enquote {\bibinfo {title} {Maintaining quantum
  coherence in the presence of noise through state monitoring},}\ }\href
  {\doibase 10.1103/PhysRevA.85.012102} {\bibfield  {journal} {\bibinfo
  {journal} {Phys. Rev. A.}\ }\textbf {\bibinfo {volume} {85}},\ \bibinfo
  {pages} {012102} (\bibinfo {year} {2012})}\BibitemShut {NoStop}%
\bibitem [{\citenamefont {Harrington}\ \emph {et~al.}(2019)\citenamefont
  {Harrington}, \citenamefont {Tan}, \citenamefont {Naghiloo},\ and\
  \citenamefont {Murch}}]{KaterMurch19}%
  \BibitemOpen
  \bibfield  {author} {\bibinfo {author} {\bibfnamefont {P.~M.}\ \bibnamefont
  {Harrington}}, \bibinfo {author} {\bibfnamefont {D.}~\bibnamefont {Tan}},
  \bibinfo {author} {\bibfnamefont {M.}~\bibnamefont {Naghiloo}}, \ and\
  \bibinfo {author} {\bibfnamefont {K.~W.}\ \bibnamefont {Murch}},\ }\bibfield
  {title} {\enquote {\bibinfo {title} {Characterizing a statistical arrow of
  time in quantum measurement dynamics},}\ }\href {\doibase
  10.1103/PhysRevLett.123.020502} {\bibfield  {journal} {\bibinfo  {journal}
  {Phys. Rev. Lett.}\ }\textbf {\bibinfo {volume} {123}},\ \bibinfo {pages}
  {020502} (\bibinfo {year} {2019})}\BibitemShut {NoStop}%
\bibitem [{\citenamefont {Wiseman}(1994)}]{Wiseman94}%
  \BibitemOpen
  \bibfield  {author} {\bibinfo {author} {\bibfnamefont {H.~M.}\ \bibnamefont
  {Wiseman}},\ }\bibfield  {title} {\enquote {\bibinfo {title} {Quantum theory
  of continuous feedback},}\ }\href {\doibase 10.1103/PhysRevA.49.2133}
  {\bibfield  {journal} {\bibinfo  {journal} {Phys. Rev. A}\ }\textbf {\bibinfo
  {volume} {49}},\ \bibinfo {pages} {2133--2150} (\bibinfo {year}
  {1994})}\BibitemShut {NoStop}%
\bibitem [{\citenamefont {Doherty}\ \emph {et~al.}(2000)\citenamefont
  {Doherty}, \citenamefont {Habib}, \citenamefont {Jacobs}, \citenamefont
  {Mabuchi},\ and\ \citenamefont {Tan}}]{Doherty_etal00}%
  \BibitemOpen
  \bibfield  {author} {\bibinfo {author} {\bibfnamefont {A.~C.}\ \bibnamefont
  {Doherty}}, \bibinfo {author} {\bibfnamefont {S.}~\bibnamefont {Habib}},
  \bibinfo {author} {\bibfnamefont {K.}~\bibnamefont {Jacobs}}, \bibinfo
  {author} {\bibfnamefont {H.}~\bibnamefont {Mabuchi}}, \ and\ \bibinfo
  {author} {\bibfnamefont {S.~M.}\ \bibnamefont {Tan}},\ }\bibfield  {title}
  {\enquote {\bibinfo {title} {Quantum feedback control and classical control
  theory},}\ }\href {\doibase 10.1103/PhysRevA.62.012105} {\bibfield  {journal}
  {\bibinfo  {journal} {Phys. Rev. A}\ }\textbf {\bibinfo {volume} {62}},\
  \bibinfo {pages} {012105} (\bibinfo {year} {2000})}\BibitemShut {NoStop}%
\bibitem [{\citenamefont {Geremia}(2006)}]{Geremia06}%
  \BibitemOpen
  \bibfield  {author} {\bibinfo {author} {\bibfnamefont {J.~M.}\ \bibnamefont
  {Geremia}},\ }\bibfield  {title} {\enquote {\bibinfo {title} {Deterministic
  and nondestructively verifiable preparation of photon number states},}\
  }\href {\doibase 10.1103/PhysRevLett.97.073601} {\bibfield  {journal}
  {\bibinfo  {journal} {Phys. Rev. Lett.}\ }\textbf {\bibinfo {volume} {97}},\
  \bibinfo {pages} {073601} (\bibinfo {year} {2006})}\BibitemShut {NoStop}%
\bibitem [{\citenamefont {Dotsenko}\ \emph {et~al.}(2009)\citenamefont
  {Dotsenko}, \citenamefont {Mirrahimi}, \citenamefont {Brune}, \citenamefont
  {Haroche}, \citenamefont {Raimond},\ and\ \citenamefont
  {Rouchon}}]{Dotsenko_etal09}%
  \BibitemOpen
  \bibfield  {author} {\bibinfo {author} {\bibfnamefont {I.}~\bibnamefont
  {Dotsenko}}, \bibinfo {author} {\bibfnamefont {M.}~\bibnamefont {Mirrahimi}},
  \bibinfo {author} {\bibfnamefont {M.}~\bibnamefont {Brune}}, \bibinfo
  {author} {\bibfnamefont {S.}~\bibnamefont {Haroche}}, \bibinfo {author}
  {\bibfnamefont {J.-M.}\ \bibnamefont {Raimond}}, \ and\ \bibinfo {author}
  {\bibfnamefont {P.}~\bibnamefont {Rouchon}},\ }\bibfield  {title} {\enquote
  {\bibinfo {title} {Quantum feedback by discrete quantum nondemolition
  measurements: Towards on-demand generation of photon-number states},}\ }\href
  {\doibase 10.1103/PhysRevA.80.013805} {\bibfield  {journal} {\bibinfo
  {journal} {Phys. Rev. A}\ }\textbf {\bibinfo {volume} {80}},\ \bibinfo
  {pages} {013805} (\bibinfo {year} {2009})}\BibitemShut {NoStop}%
\bibitem [{\citenamefont {Gillett}\ \emph {et~al.}(2010)\citenamefont
  {Gillett}, \citenamefont {Dalton}, \citenamefont {Lanyon}, \citenamefont
  {Almeida}, \citenamefont {Barbieri}, \citenamefont {Pryde}, \citenamefont
  {O'Brien}, \citenamefont {Resch}, \citenamefont {Bartlett},\ and\
  \citenamefont {White}}]{Gillett_etal10}%
  \BibitemOpen
  \bibfield  {author} {\bibinfo {author} {\bibfnamefont {G.~G.}\ \bibnamefont
  {Gillett}}, \bibinfo {author} {\bibfnamefont {R.~B.}\ \bibnamefont {Dalton}},
  \bibinfo {author} {\bibfnamefont {B.~P.}\ \bibnamefont {Lanyon}}, \bibinfo
  {author} {\bibfnamefont {M.~P.}\ \bibnamefont {Almeida}}, \bibinfo {author}
  {\bibfnamefont {M.}~\bibnamefont {Barbieri}}, \bibinfo {author}
  {\bibfnamefont {G.~J.}\ \bibnamefont {Pryde}}, \bibinfo {author}
  {\bibfnamefont {J.~L.}\ \bibnamefont {O'Brien}}, \bibinfo {author}
  {\bibfnamefont {K.~J.}\ \bibnamefont {Resch}}, \bibinfo {author}
  {\bibfnamefont {S.~D.}\ \bibnamefont {Bartlett}}, \ and\ \bibinfo {author}
  {\bibfnamefont {A.~G.}\ \bibnamefont {White}},\ }\bibfield  {title} {\enquote
  {\bibinfo {title} {Experimental feedback control of quantum systems using
  weak measurements},}\ }\href {\doibase 10.1103/PhysRevLett.104.080503}
  {\bibfield  {journal} {\bibinfo  {journal} {Phys. Rev. Lett.}\ }\textbf
  {\bibinfo {volume} {104}},\ \bibinfo {pages} {080503} (\bibinfo {year}
  {2010})}\BibitemShut {NoStop}%
\bibitem [{\citenamefont {Sayrin}\ \emph {et~al.}(2011)\citenamefont {Sayrin},
  \citenamefont {Dotsenko}, \citenamefont {Zhou}, \citenamefont {Peaudecerf},
  \citenamefont {Rybarczyk}, \citenamefont {Gleyzes}, \citenamefont {Rouchon},
  \citenamefont {Mirrahimi}, \citenamefont {Amini}, \citenamefont {Brune},
  \citenamefont {Raimond},\ and\ \citenamefont {Haroche}}]{Sayrin2011}%
  \BibitemOpen
  \bibfield  {author} {\bibinfo {author} {\bibfnamefont {C.}~\bibnamefont
  {Sayrin}}, \bibinfo {author} {\bibfnamefont {I.}~\bibnamefont {Dotsenko}},
  \bibinfo {author} {\bibfnamefont {X.}~\bibnamefont {Zhou}}, \bibinfo {author}
  {\bibfnamefont {B.}~\bibnamefont {Peaudecerf}}, \bibinfo {author}
  {\bibfnamefont {T.}~\bibnamefont {Rybarczyk}}, \bibinfo {author}
  {\bibfnamefont {S.}~\bibnamefont {Gleyzes}}, \bibinfo {author} {\bibfnamefont
  {P.}~\bibnamefont {Rouchon}}, \bibinfo {author} {\bibfnamefont
  {M.}~\bibnamefont {Mirrahimi}}, \bibinfo {author} {\bibfnamefont
  {H.}~\bibnamefont {Amini}}, \bibinfo {author} {\bibfnamefont
  {M.}~\bibnamefont {Brune}}, \bibinfo {author} {\bibfnamefont {J.-M.}\
  \bibnamefont {Raimond}}, \ and\ \bibinfo {author} {\bibfnamefont
  {S.}~\bibnamefont {Haroche}},\ }\bibfield  {title} {\enquote {\bibinfo
  {title} {Real-time quantum feedback prepares and stabilizes photon number
  states},}\ }\href {\doibase 10.1038/nature10376} {\bibfield  {journal}
  {\bibinfo  {journal} {Nature}\ }\textbf {\bibinfo {volume} {477}},\ \bibinfo
  {pages} {73--77} (\bibinfo {year} {2011})}\BibitemShut {NoStop}%
\bibitem [{\citenamefont {Vijay}\ \emph {et~al.}(2012)\citenamefont {Vijay},
  \citenamefont {Macklin}, \citenamefont {Slichter}, \citenamefont {Weber},
  \citenamefont {Murch}, \citenamefont {Naik}, \citenamefont {Korotkov},\ and\
  \citenamefont {Siddiqi}}]{Vijay2012}%
  \BibitemOpen
  \bibfield  {author} {\bibinfo {author} {\bibfnamefont {R.}~\bibnamefont
  {Vijay}}, \bibinfo {author} {\bibfnamefont {C.}~\bibnamefont {Macklin}},
  \bibinfo {author} {\bibfnamefont {D.~H.}\ \bibnamefont {Slichter}}, \bibinfo
  {author} {\bibfnamefont {S.~J.}\ \bibnamefont {Weber}}, \bibinfo {author}
  {\bibfnamefont {K.~W.}\ \bibnamefont {Murch}}, \bibinfo {author}
  {\bibfnamefont {R.}~\bibnamefont {Naik}}, \bibinfo {author} {\bibfnamefont
  {A.~N.}\ \bibnamefont {Korotkov}}, \ and\ \bibinfo {author} {\bibfnamefont
  {I.}~\bibnamefont {Siddiqi}},\ }\bibfield  {title} {\enquote {\bibinfo
  {title} {Stabilizing {R}abi oscillations in a superconducting qubit using
  quantum feedback},}\ }\href {https://doi.org/10.1038/nature11505} {\bibfield
  {journal} {\bibinfo  {journal} {Nature}\ }\textbf {\bibinfo {volume} {490}},\
  \bibinfo {pages} {77--80} (\bibinfo {year} {2012})}\BibitemShut {NoStop}%
\bibitem [{\citenamefont {Lloyd}(2000)}]{Lloyd00}%
  \BibitemOpen
  \bibfield  {author} {\bibinfo {author} {\bibfnamefont {S.}~\bibnamefont
  {Lloyd}},\ }\bibfield  {title} {\enquote {\bibinfo {title} {Coherent quantum
  feedback},}\ }\href {\doibase 10.1103/PhysRevA.62.022108} {\bibfield
  {journal} {\bibinfo  {journal} {Phys. Rev. A}\ }\textbf {\bibinfo {volume}
  {62}},\ \bibinfo {pages} {022108} (\bibinfo {year} {2000})}\BibitemShut
  {NoStop}%
\bibitem [{\citenamefont {Nelson}\ \emph {et~al.}(2000)\citenamefont {Nelson},
  \citenamefont {Weinstein}, \citenamefont {Cory},\ and\ \citenamefont
  {Lloyd}}]{Nelson_etal00}%
  \BibitemOpen
  \bibfield  {author} {\bibinfo {author} {\bibfnamefont {R.~J.}\ \bibnamefont
  {Nelson}}, \bibinfo {author} {\bibfnamefont {Y.}~\bibnamefont {Weinstein}},
  \bibinfo {author} {\bibfnamefont {D.}~\bibnamefont {Cory}}, \ and\ \bibinfo
  {author} {\bibfnamefont {S.}~\bibnamefont {Lloyd}},\ }\bibfield  {title}
  {\enquote {\bibinfo {title} {Experimental demonstration of fully coherent
  quantum feedback},}\ }\href {\doibase 10.1103/PhysRevLett.85.3045} {\bibfield
   {journal} {\bibinfo  {journal} {Phys. Rev. Lett.}\ }\textbf {\bibinfo
  {volume} {85}},\ \bibinfo {pages} {3045--3048} (\bibinfo {year}
  {2000})}\BibitemShut {NoStop}%
\bibitem [{\citenamefont {Hirose}\ and\ \citenamefont
  {Cappellaro}(2016)}]{HiroseCappellaro16}%
  \BibitemOpen
  \bibfield  {author} {\bibinfo {author} {\bibfnamefont {M.}~\bibnamefont
  {Hirose}}\ and\ \bibinfo {author} {\bibfnamefont {P.}~\bibnamefont
  {Cappellaro}},\ }\bibfield  {title} {\enquote {\bibinfo {title} {Coherent
  feedback control of a single qubit in diamond},}\ }\href {\doibase
  10.1038/nature17404} {\bibfield  {journal} {\bibinfo  {journal} {Nature}\
  }\textbf {\bibinfo {volume} {532}},\ \bibinfo {pages} {77--80} (\bibinfo
  {year} {2016})}\BibitemShut {NoStop}%
\bibitem [{\citenamefont {Hamerly}\ and\ \citenamefont
  {Mabuchi}(2012)}]{HamerlyMabuchi12}%
  \BibitemOpen
  \bibfield  {author} {\bibinfo {author} {\bibfnamefont {R.}~\bibnamefont
  {Hamerly}}\ and\ \bibinfo {author} {\bibfnamefont {H.}~\bibnamefont
  {Mabuchi}},\ }\bibfield  {title} {\enquote {\bibinfo {title} {Advantages of
  coherent feedback for cooling quantum oscillators},}\ }\href {\doibase
  10.1103/PhysRevLett.109.173602} {\bibfield  {journal} {\bibinfo  {journal}
  {Phys. Rev. Lett.}\ }\textbf {\bibinfo {volume} {109}},\ \bibinfo {pages}
  {173602} (\bibinfo {year} {2012})}\BibitemShut {NoStop}%
\bibitem [{\citenamefont {Yang}\ \emph {et~al.}(2015)\citenamefont {Yang},
  \citenamefont {Zhang}, \citenamefont {Wang}, \citenamefont {Liu},
  \citenamefont {Wu}, \citenamefont {Liu}, \citenamefont {Li},\ and\
  \citenamefont {Nori}}]{Yang_etal15}%
  \BibitemOpen
  \bibfield  {author} {\bibinfo {author} {\bibfnamefont {N.}~\bibnamefont
  {Yang}}, \bibinfo {author} {\bibfnamefont {J.}~\bibnamefont {Zhang}},
  \bibinfo {author} {\bibfnamefont {H.}~\bibnamefont {Wang}}, \bibinfo {author}
  {\bibfnamefont {Y.-X.}\ \bibnamefont {Liu}}, \bibinfo {author} {\bibfnamefont
  {R.-B.}\ \bibnamefont {Wu}}, \bibinfo {author} {\bibfnamefont {L.-Q}\
  \bibnamefont {Liu}}, \bibinfo {author} {\bibfnamefont {C.-W.}\ \bibnamefont
  {Li}}, \ and\ \bibinfo {author} {\bibfnamefont {F.}~\bibnamefont {Nori}},\
  }\bibfield  {title} {\enquote {\bibinfo {title} {Noise suppression of on-chip
  mechanical resonators by chaotic coherent feedback},}\ }\href {\doibase
  10.1103/PhysRevA.92.033812} {\bibfield  {journal} {\bibinfo  {journal} {Phys.
  Rev. A}\ }\textbf {\bibinfo {volume} {92}},\ \bibinfo {pages} {033812}
  (\bibinfo {year} {2015})}\BibitemShut {NoStop}%
\bibitem [{\citenamefont {Hein}\ \emph {et~al.}(2015)\citenamefont {Hein},
  \citenamefont {Schulze}, \citenamefont {Carmele},\ and\ \citenamefont
  {Knorr}}]{Hein_etal15}%
  \BibitemOpen
  \bibfield  {author} {\bibinfo {author} {\bibfnamefont {S.~M.}\ \bibnamefont
  {Hein}}, \bibinfo {author} {\bibfnamefont {F.}~\bibnamefont {Schulze}},
  \bibinfo {author} {\bibfnamefont {A.}~\bibnamefont {Carmele}}, \ and\
  \bibinfo {author} {\bibfnamefont {A.}~\bibnamefont {Knorr}},\ }\bibfield
  {title} {\enquote {\bibinfo {title} {Entanglement control in quantum networks
  by quantum-coherent time-delayed feedback},}\ }\href {\doibase
  10.1103/PhysRevA.91.052321} {\bibfield  {journal} {\bibinfo  {journal} {Phys.
  Rev. A}\ }\textbf {\bibinfo {volume} {91}},\ \bibinfo {pages} {052321}
  (\bibinfo {year} {2015})}\BibitemShut {NoStop}%
\bibitem [{\citenamefont {Uys}\ \emph {et~al.}(2018)\citenamefont {Uys},
  \citenamefont {Bassa}, \citenamefont {du~Toit}, \citenamefont {Ghosh},\ and\
  \citenamefont {Konrad}}]{Uys2018}%
  \BibitemOpen
  \bibfield  {author} {\bibinfo {author} {\bibfnamefont {H.}~\bibnamefont
  {Uys}}, \bibinfo {author} {\bibfnamefont {H.}~\bibnamefont {Bassa}}, \bibinfo
  {author} {\bibfnamefont {P.}~\bibnamefont {du~Toit}}, \bibinfo {author}
  {\bibfnamefont {S.}~\bibnamefont {Ghosh}}, \ and\ \bibinfo {author}
  {\bibfnamefont {T.}~\bibnamefont {Konrad}},\ }\bibfield  {title} {\enquote
  {\bibinfo {title} {Quantum control through measurement feedback},}\ }\href
  {\doibase 10.1103/PhysRevA.97.060102} {\bibfield  {journal} {\bibinfo
  {journal} {Phys. Rev. A}\ }\textbf {\bibinfo {volume} {97}},\ \bibinfo
  {pages} {060102} (\bibinfo {year} {2018})}\BibitemShut {NoStop}%
\bibitem [{\citenamefont {Kraus}(1983)}]{Kraus83}%
  \BibitemOpen
  \bibfield  {author} {\bibinfo {author} {\bibfnamefont {K.}~\bibnamefont
  {Kraus}},\ }\href@noop {} {\emph {\bibinfo {title} {States, Effects, and
  Operations}}}\ (\bibinfo  {publisher} {Springer-Verlag},\ \bibinfo {address}
  {Berlin},\ \bibinfo {year} {1983})\BibitemShut {NoStop}%
\bibitem [{\citenamefont {Imamoglu}\ \emph {et~al.}(1999)\citenamefont
  {Imamoglu}, \citenamefont {Awschalom}, \citenamefont {Burkard}, \citenamefont
  {DiVincenzo}, \citenamefont {Loss}, \citenamefont {Sherwin},\ and\
  \citenamefont {Small}}]{Imamoglu99}%
  \BibitemOpen
  \bibfield  {author} {\bibinfo {author} {\bibfnamefont {A.}~\bibnamefont
  {Imamoglu}}, \bibinfo {author} {\bibfnamefont {D.~D.}\ \bibnamefont
  {Awschalom}}, \bibinfo {author} {\bibfnamefont {G.}~\bibnamefont {Burkard}},
  \bibinfo {author} {\bibfnamefont {D.~P.}\ \bibnamefont {DiVincenzo}},
  \bibinfo {author} {\bibfnamefont {D.}~\bibnamefont {Loss}}, \bibinfo {author}
  {\bibfnamefont {M.}~\bibnamefont {Sherwin}}, \ and\ \bibinfo {author}
  {\bibfnamefont {A.}~\bibnamefont {Small}},\ }\bibfield  {title} {\enquote
  {\bibinfo {title} {Quantum information processing using quantum dot spins and
  cavity {QED}},}\ }\href {\doibase 10.1103/PhysRevLett.83.4204} {\bibfield
  {journal} {\bibinfo  {journal} {Phys. Rev. Lett.}\ }\textbf {\bibinfo
  {volume} {83}},\ \bibinfo {pages} {4204--4207} (\bibinfo {year}
  {1999})}\BibitemShut {NoStop}%
\bibitem [{\citenamefont {Siewert}\ \emph {et~al.}(2000)\citenamefont
  {Siewert}, \citenamefont {Fazio}, \citenamefont {Palma},\ and\ \citenamefont
  {Sciacca}}]{Siewert00}%
  \BibitemOpen
  \bibfield  {author} {\bibinfo {author} {\bibfnamefont {J.}~\bibnamefont
  {Siewert}}, \bibinfo {author} {\bibfnamefont {R.}~\bibnamefont {Fazio}},
  \bibinfo {author} {\bibfnamefont {G.~M.}\ \bibnamefont {Palma}}, \ and\
  \bibinfo {author} {\bibfnamefont {E.}~\bibnamefont {Sciacca}},\ }\bibfield
  {title} {\enquote {\bibinfo {title} {Aspects of qubit dynamics in the
  presence of leakage},}\ }\href {\doibase 10.1023/A:1004612016347} {\bibfield
  {journal} {\bibinfo  {journal} {J. Low Temp. Phys.}\ }\textbf {\bibinfo
  {volume} {118}},\ \bibinfo {pages} {795--804} (\bibinfo {year}
  {2000})}\BibitemShut {NoStop}%
\bibitem [{\citenamefont {Mozyrsky}\ \emph {et~al.}(2001)\citenamefont
  {Mozyrsky}, \citenamefont {Privman},\ and\ \citenamefont
  {Glasser}}]{Lawrence01}%
  \BibitemOpen
  \bibfield  {author} {\bibinfo {author} {\bibfnamefont {D.}~\bibnamefont
  {Mozyrsky}}, \bibinfo {author} {\bibfnamefont {V.}~\bibnamefont {Privman}}, \
  and\ \bibinfo {author} {\bibfnamefont {M.~L.}\ \bibnamefont {Glasser}},\
  }\bibfield  {title} {\enquote {\bibinfo {title} {Indirect interaction of
  solid-state qubits via two-dimensional electron gas},}\ }\href {\doibase
  10.1103/PhysRevLett.86.5112} {\bibfield  {journal} {\bibinfo  {journal}
  {Phys. Rev. Lett.}\ }\textbf {\bibinfo {volume} {86}},\ \bibinfo {pages}
  {5112--5115} (\bibinfo {year} {2001})}\BibitemShut {NoStop}%
\bibitem [{\citenamefont {Loss}\ and\ \citenamefont
  {DiVincenzo}(1998)}]{Loss98}%
  \BibitemOpen
  \bibfield  {author} {\bibinfo {author} {\bibfnamefont {D.}~\bibnamefont
  {Loss}}\ and\ \bibinfo {author} {\bibfnamefont {D.~P.}\ \bibnamefont
  {DiVincenzo}},\ }\bibfield  {title} {\enquote {\bibinfo {title} {Quantum
  computation with quantum dots},}\ }\href {\doibase 10.1103/PhysRevA.57.120}
  {\bibfield  {journal} {\bibinfo  {journal} {Phys. Rev. A}\ }\textbf {\bibinfo
  {volume} {57}},\ \bibinfo {pages} {120--126} (\bibinfo {year}
  {1998})}\BibitemShut {NoStop}%
\bibitem [{\citenamefont {Kane}(1998)}]{Kane98}%
  \BibitemOpen
  \bibfield  {author} {\bibinfo {author} {\bibfnamefont {B.~E.}\ \bibnamefont
  {Kane}},\ }\bibfield  {title} {\enquote {\bibinfo {title} {A silicon-based
  nuclear spin quantum computer},}\ }\href {\doibase 10.1038/30156} {\bibfield
  {journal} {\bibinfo  {journal} {Nature}\ }\textbf {\bibinfo {volume} {393}},\
  \bibinfo {pages} {133--137} (\bibinfo {year} {1998})}\BibitemShut {NoStop}%
\bibitem [{\citenamefont {Vrijen}\ \emph {et~al.}(2000)\citenamefont {Vrijen},
  \citenamefont {Yablonovitch}, \citenamefont {Wang}, \citenamefont {Jiang},
  \citenamefont {Balandin}, \citenamefont {Roychowdhury}, \citenamefont {Mor},\
  and\ \citenamefont {DiVincenzo}}]{Vrijen00}%
  \BibitemOpen
  \bibfield  {author} {\bibinfo {author} {\bibfnamefont {R.}~\bibnamefont
  {Vrijen}}, \bibinfo {author} {\bibfnamefont {E.}~\bibnamefont
  {Yablonovitch}}, \bibinfo {author} {\bibfnamefont {K.}~\bibnamefont {Wang}},
  \bibinfo {author} {\bibfnamefont {H.~W.}\ \bibnamefont {Jiang}}, \bibinfo
  {author} {\bibfnamefont {A.}~\bibnamefont {Balandin}}, \bibinfo {author}
  {\bibfnamefont {V.}~\bibnamefont {Roychowdhury}}, \bibinfo {author}
  {\bibfnamefont {T.}~\bibnamefont {Mor}}, \ and\ \bibinfo {author}
  {\bibfnamefont {D.}~\bibnamefont {DiVincenzo}},\ }\bibfield  {title}
  {\enquote {\bibinfo {title} {Electron-spin-resonance transistors for quantum
  computing in silicon-germanium heterostructures},}\ }\href {\doibase
  10.1103/PhysRevA.62.012306} {\bibfield  {journal} {\bibinfo  {journal} {Phys.
  Rev. A}\ }\textbf {\bibinfo {volume} {62}},\ \bibinfo {pages} {012306}
  (\bibinfo {year} {2000})}\BibitemShut {NoStop}%
\bibitem [{\citenamefont {Plenio}\ and\ \citenamefont
  {Virmani}(2014)}]{Plenio2014}%
  \BibitemOpen
  \bibfield  {author} {\bibinfo {author} {\bibfnamefont {M.~B.}\ \bibnamefont
  {Plenio}}\ and\ \bibinfo {author} {\bibfnamefont {S.~S.}\ \bibnamefont
  {Virmani}},\ }\enquote {\bibinfo {title} {An introduction to entanglement
  theory},}\ in\ \href {\doibase 10.1007/978-3-319-04063-9_8} {\emph {\bibinfo
  {booktitle} {Quantum Information and Coherence}}},\ \bibinfo {editor} {edited
  by\ \bibinfo {editor} {\bibfnamefont {E.}~\bibnamefont {Andersson}}\ and\
  \bibinfo {editor} {\bibfnamefont {P.}~\bibnamefont {{\"O}hberg}}}\ (\bibinfo
  {publisher} {Springer},\ \bibinfo {address} {Cham},\ \bibinfo {year} {2014})\
  pp.\ \bibinfo {pages} {173--209}\BibitemShut {NoStop}%
\end{thebibliography}%


\begin{thebibliography}{3}%
\makeatletter
\providecommand \@ifxundefined [1]{%
 \@ifx{#1\undefined}
}%
\providecommand \@ifnum [1]{%
 \ifnum #1\expandafter \@firstoftwo
 \else \expandafter \@secondoftwo
 \fi
}%
\providecommand \@ifx [1]{%
 \ifx #1\expandafter \@firstoftwo
 \else \expandafter \@secondoftwo
 \fi
}%
\providecommand \natexlab [1]{#1}%
\providecommand \enquote  [1]{``#1''}%
\providecommand \bibnamefont  [1]{#1}%
\providecommand \bibfnamefont [1]{#1}%
\providecommand \citenamefont [1]{#1}%
\providecommand \href@noop [0]{\@secondoftwo}%
\providecommand \href [0]{\begingroup \@sanitize@url \@href}%
\providecommand \@href[1]{\@@startlink{#1}\@@href}%
\providecommand \@@href[1]{\endgroup#1\@@endlink}%
\providecommand \@sanitize@url [0]{\catcode `\\12\catcode `\$12\catcode
  `\&12\catcode `\#12\catcode `\^12\catcode `\_12\catcode `\%12\relax}%
\providecommand \@@startlink[1]{}%
\providecommand \@@endlink[0]{}%
\providecommand \url  [0]{\begingroup\@sanitize@url \@url }%
\providecommand \@url [1]{\endgroup\@href {#1}{\urlprefix }}%
\providecommand \urlprefix  [0]{URL }%
\providecommand \Eprint [0]{\href }%
\providecommand \doibase [0]{http://dx.doi.org/}%
\providecommand \selectlanguage [0]{\@gobble}%
\providecommand \bibinfo  [0]{\@secondoftwo}%
\providecommand \bibfield  [0]{\@secondoftwo}%
\providecommand \translation [1]{[#1]}%
\providecommand \BibitemOpen [0]{}%
\providecommand \bibitemStop [0]{}%
\providecommand \bibitemNoStop [0]{.\EOS\space}%
\providecommand \EOS [0]{\spacefactor3000\relax}%
\providecommand \BibitemShut  [1]{\csname bibitem#1\endcsname}%
\let\auto@bib@innerbib\@empty
\bibitem [{\citenamefont {Barchielli}\ \emph {et~al.}(1982)\citenamefont
  {Barchielli}, \citenamefont {Lanz},\ and\ \citenamefont
  {Prosperi}}]{Barchielli.et.al82}%
  \BibitemOpen
  \bibfield  {author} {\bibinfo {author} {\bibfnamefont {A..}\ \bibnamefont
  {Barchielli}}, \bibinfo {author} {\bibfnamefont {L.}~\bibnamefont {Lanz}}, \
  and\ \bibinfo {author} {\bibfnamefont {G.M.}\ \bibnamefont {Prosperi}},\
  }\bibfield  {title} {\enquote {\bibinfo {title} {A model for the macroscopic
  description and continual observation in quantum mechanics.}}\ }\href@noop {}
  {\bibfield  {journal} {\bibinfo  {journal} {Nuovo Cimento}\ }\textbf
  {\bibinfo {volume} {B72}},\ \bibinfo {pages} {79} (\bibinfo {year}
  {1982})}\BibitemShut {NoStop}%
\bibitem [{\citenamefont {Audretsch}\ \emph {et~al.}(2002)\citenamefont
  {Audretsch}, \citenamefont {Di{\'o}si},\ and\ \citenamefont
  {Konrad}}]{AudretschDiosiKonrad02}%
  \BibitemOpen
  \bibfield  {author} {\bibinfo {author} {\bibfnamefont {J.}~\bibnamefont
  {Audretsch}}, \bibinfo {author} {\bibfnamefont {L.}~\bibnamefont
  {Di{\'o}si}}, \ and\ \bibinfo {author} {\bibfnamefont {T.}~\bibnamefont
  {Konrad}},\ }\bibfield  {title} {\enquote {\bibinfo {title} {Evolution of a
  qubit under the influence of a succession of weak measurements},}\
  }\href@noop {} {\bibfield  {journal} {\bibinfo  {journal} {Phys. Rev. A}\
  }\textbf {\bibinfo {volume} {66}},\ \bibinfo {pages} {022310} (\bibinfo
  {year} {2002})}\BibitemShut {NoStop}%
\bibitem [{\citenamefont {Elaydi}(2005)}]{Elaydi}%
  \BibitemOpen
  \bibfield  {author} {\bibinfo {author} {\bibfnamefont {S.}~\bibnamefont
  {Elaydi}},\ }\href {\doibase 10.1007/0-387-27602-5} {\emph {\bibinfo {title}
  {An Introduction to Difference Equations}}}\ (\bibinfo  {publisher}
  {Springer-Verlag},\ \bibinfo {address} {New York},\ \bibinfo {year}
  {2005})\BibitemShut {NoStop}%
\end{thebibliography}%

\section*{Acknowledgements}
A.R. acknowledges financial support from the National Research Foundation of South Africa. We thank Pamela Benporath for the design of Figs.~1 and 4.

\section*{Author contributions}
The conceptual idea was formulated by T.K.\ and H.U.\ with input from M.K.. The manuscript was written by T.K.\  and M.K.\ with input from A.R.\   Proof of convergence and construction methods were laid out by T.K.. 
Examples 1, 2 and 3 were formulated by T.K., A.R., and M.K.\ respectively.
Simulations where performed by A.R..   
The supplementary information was written by A.R.\ with input from T.K. and M.K.. 
All authors contributed to the editing of the manuscript.

\section*{Competing financial interests}
The authors declare no competing financial interests.


\end{document}


\title{Supplementary Information}
%
\maketitle

\appendix
\renewcommand{\appendixname}{}

\section{Construction of Kraus operators}
Here we prove that the second construction method {in Sec.~II.C of the main text} leads to a set of Kraus operators which satisfy the conditions for convergence to the target state, Eqs.\ (6) and (14) in the main text. 
\begin{lem}
Given a POVM \AR{$(E_i)_{i \in 0,\dots, d-1}$} with pairwise commuting effects $E_i$, we denote by \AR{$(\ket{k})_{k=0,\dots,d-1}$} the simultaneous eigenstates of the $E_i$. Assume that the POVM corresponds to an unsharp measurement of a nondegenerate observable, {i.e.,\AR{ $E_i= \sum_{k=0}^{d-1} \lambda_{ik} \pro{k}$} with $\det(\lambda_{ik}) \not=0$}. For a target state \AR{$\ket{T} = \sum_{k=0}^{m-1} t_k \ket{k}$} with $m<d$ nonzero coefficients $t_k$, we define a {modified} state \AR{$\ket{T'} \propto \ket{T} + \sum_{k=m}^{d -1}t_k \ket{k}$} with arbitrary nonzero coefficients \AR{$(t_k)_{k=m,\dots, d-1}$}. Furthermore, assume $E_i \ket{T} \neq 0$ for exactly $m$ of the effects $E_i$. Then it follows that
\AR{\begin{align}\label{e:span}
\Span(E_0\ket{T}, \dots, E_{m-1} \ket{T}, E_{m}\ket{T'},\dots, E_{d-1} \ket{T'})= \mathcal{H}.
\end{align}}
\end{lem}

Proof: Without loss of generality we label the effects such that $E_i\ket{T}\neq 0$ for $i=1,\dots, m$. Since {all expansion coefficients $t_k$ of $\ket{T'}$ are nonzero,} \AR{$(E_i\ket{T'})_{i=0,\dots, d-1}$} spans the full Hilbert space. \ar{This can be seen from
\begin{align}
\det(\lambda_{ik}t_k) = \det(\lambda_{ik})\AR{\prod_{k=0}^{d-1} t_k}\,.
\end{align}
}
Given that \AR{$E_i \ket{T} = \sum_{k=0}^{m-1} \lambda_{ik} t_k \ket{k} = 0$} for \AR{$i = m,\dots,d-1$}, {it follows that} $\lambda_{ik} = 0$ for \AR{$i=m,\dots,d-1$} and \AR{$k=0,\dots,m-1$}. The determinant can now be written as
\begin{align}
\det(\lambda_{ik})  = \det(\lambda_{ik})_{\AR{ i,k=0,\dots, m-1}} \det(\lambda_{ik})_{\AR{i,k=m,\dots, d-1}} \neq 0,
\end{align}
and consequently $\AR{\Span(E_i \ket{T'})_{i=m,\dots,d-1}} = \mathcal{H}_{d-m}$ with $ \mathcal{H}_m \oplus \mathcal{H}_{d-m} = \mathcal{H}$. \qed

{Using Eq.~\eqref{e:span},} the Kraus operators $M_i = U_i \sqrt{E_i}$ with unitary $U_i$ satisfying $U_i \sqrt{E_i} \ket{T} \propto \ket{T}$ for $\sqrt{E}_i \ket{T} \neq 0$ and $U_i \sqrt{E_i} \ket{T'}\propto \ket{T}$ otherwise, can easily be shown to satisfy Eqs.~(6) and (14). 

\section{Exponential increase of target fidelity for qubits}
In this section we study the target fidelity of a sequence of control operations that satisfy the conditions for convergence to the target state, Eqs.\ (6) and (14) of the main text. For systems with a two-dimensional Hilbert space there is only one state $\ket{T_\perp}$ orthogonal to the target state $\ket{T}$, and thus 
\begin{equation}
\pro{T} + \pro{T_\perp}= \mathbb{I}\,. 
\label{completeness}
\end{equation}
The change in fidelity, Eqn.~(10), then reads
\begin{align}
\Delta F(\rho) & =  \sum_i \bigl\lVert \sqrt{\rho}  \ket{T_\perp}\braket{T_\perp|M_i^\dagger|T}\bigr\rVert^2 \nonumber \\
& =  \sum_i  \bigl\lvert\braket{T_\perp|M_i^\dagger|T}\bigr\rvert^2  \braket{T_\perp|\rho|T_\perp} \nonumber \\
& = \gamma [1- F(\rho,T)] \label{eq:delta_F},
\end{align}
where we used (\ref{completeness}) to arrive at $\braket{T_\perp|\rho|T_\perp} =  1 -  \braket{T|\rho|T} = [1- {F(\rho,T)}]$. The constant $\gamma$, which represents the control strength, can be evaluated based on Eq.~\eqref{completeness} as well as  Eqs.~(6) and (7) as follows, 
\begin{align}
\gamma &\equiv \sum_i \bigl\lVert\braket{T_\perp|M_i^\dagger|T}\bigr\rVert^2 \nonumber \\
&= \sum_i \braket{T|M_i(\mathbb{I} -  \pro{T}) M_i^\dagger|T}\nonumber \\
& = \sum_i \left(\braket{T|M_i M_i^\dagger|T} - \braket{T|M_i|T}\braket{T|M_i^\dagger |T}\right)  \nonumber\\
& = \sum_i \braket{T|M_i M_i^\dagger|T} - 1.\label{gamma}
\end{align}

In order to find the solution to the difference equation (\ref{eq:delta_F}), we derive a differential equation by introducing the continuum limit of sequential control operations. 
For this purpose, we consider a sequence of control operations with vanishing time period between consecutive operations, $\Delta t \rightarrow 0$. This requires  {one} to consider also infinitely weak control operations, i.e. a vanishing control strength,  $\gamma\rightarrow 0$, in order to obtain a finite rate $\Gamma$ of fidelity increase  for continuous control     
\begin{align}
\lim_{\substack{\Delta t \rightarrow 0\\ \gamma \rightarrow 0}}  \frac{\gamma}{\Delta t} = \Gamma \label{eq:limit}\,.
\end{align}
This limit is reminiscent of the continuum limit of weak measurements \cite{Barchielli.et.al82,  AudretschDiosiKonrad02}, {which} connects sequential with continuous measurement.
{In this way} we obtain the following differential equation for the fidelity as the continuum limit of the difference equation for a single step \eqref{eq:delta_F},
\begin{align}
\frac{dF}{dt} = \lim_{\substack{\Delta t \rightarrow 0\\ \gamma \rightarrow 0}}  \frac{\Delta F}{\Delta t} = \Gamma (1-F).
\label{eq:dF}
\end{align}
The target fidelity as function of time is thus given by
\begin{align}
F(t) = 1- (1-F_0) e^{-\Gamma t} \label{eq:F},
\end{align}
where $F_0 =  {F(\rho,T)}$ is the initial value of the overlap with the target state. 
Alternatively, the general solution $F(t)$ of the inhomogeneous linear difference equation \eqref{eq:delta_F} can be proved by induction \cite[Eqs.(1.2.6) \& (1.2.8)]{Elaydi}.

In order to express the rate of fidelity increase $\Gamma$ in terms of its discrete analog $\gamma$ we consider the increase of the target fidelity after a single control operation, 
\begin{align}
\Delta F= F(\Delta t) - F_0 = (1- F_0) (1-e^{-\Gamma \Delta t}),
\end{align}
and insert the expression for $\Delta F$ from Eq.\ \eqref{eq:delta_F}. Measuring time in units of $\Delta t$, we thus obtain  
\begin{align}
1 -\gamma & = e^{-\Gamma}.
\end{align}
Hence, the expression for the target fidelity (\ref{eq:F})  after  $n$ discrete control operations reads   
\begin{align}
F( {\$^n(\rho),T}) = 1- (1-F_0) (1-\gamma)^n.
\end{align}

\section{Exponential increase of the target fidelity for weak swap channel}
The target fidelity of a system with  {$d$-dimensional} Hilbert space and weak swap interaction  {(23)} resulting in channel  {(24)}  increases exponentially with the number of control operations. The Kraus operators  {(24)} lead to a change in fidelity (10) of
\begin{align}
\Delta F(\rho) & =\AR{ \sum_{i=0}^{d-1} }\bigl\lVert \sqrt{\rho} (\mathbb{I}-\ket{T}\bra{T}) M_i^\dagger \ket{T} \bigr\rVert^2 \nonumber\\
& = 0 +\AR{ \sum_{i=1}^{d-1} }\sin^2\lambda \braket{\psi_i|\rho|\psi_i} \nonumber\\
& = \sin^2(\lambda) \left[1-  {F(\rho,T)}\right]. \label{eq:delta_F_new}
\end{align}
By the same arguments as in the  previous section, we obtain the expression for the target fidelity after $n$ discrete control operations
\begin{align}
F( {\$^n(\rho),T}) = 1- (1-F_0) [1 - \sin^2(\lambda)]^n,
\end{align}
where $F_0{=F(\rho,T)}$ is the initial fidelity.

\section{Decomposition of swap into qubit swaps}
The swap operator with respect to the system basis \AR{$(\ket{k})_{k=0,1,\ldots, d-1}$} and the controller basis \AR{$(\ket{k}_c)_{k=0,1,\ldots, n-1}$} reads
\begin{align}\label{e:swap}
S=\AR{\sum_{l,k =0}^{d-1} }\ket{k}\bra{l}\otimes  \ket{l}_c{\vphantom{\ket{l}}}_c\!\bra{k},
\end{align} 
where we assume that $d=n$. Note that the swap operator is form-invariant under a basis change given by the unitary transformation $U\otimes U$. If the Hilbert space dimensions are powers of two, $d=2^m$, then each basis vector can be relabelled using the binary representation of the integer $k$, $\ket{k}\equiv\ket{k_{1} k_{2} \dots k_{m}}$ with \AR{$k_{i} \in \{0,1\}$} for $i = 1, 2 \dots m$ (and similarly for the controller states). In this case the swap operator becomes
\begin{align}
S & =\AR{\sum_{\substack{l_1 \dots l_m =0 \\ k_1 \dots k_m=0}}^{1}} \ket{k_1 \dots k_m}\bra{l_1 \dots l_m}\otimes  \ket{l_1 \dots l_m}_c{\vphantom{\ket{l_1 \dots l_m}}}_c\!\bra{k_1 \dots k_m}\nonumber\\
& = \AR{\bigotimes_{i=0}^{m-1}} S_i,
\end{align}
where 
\begin{equation}
S_i=\AR{\sum_{l_i, k_i = 0}^1} \ket{k_i}\bra{l_i}\otimes  \ket{l_i}_c{\vphantom{\ket{l_i}}}_c\!\bra{k_i}
\end{equation} 
is the operator which swaps the $i$th qubit of the system with the $i$th qubit of the controller. Implementing the control channel  {(24)} by a unitary $U=\exp(-i \lambda S)$ based on qubit swaps therefore requires a Hamiltonian $H\propto S$
that contains  {not only pair interactions, but} interactions between $m$ qubits, which most likely makes the experimental implementation of the required unitaries very challenging. As a potential way around this difficulty we explore in the following whether {pairwise interactions between system and controller qubits} can generate a satisfactory approximate control channel.

\section{Approximate control by pairwise interactions}
We consider the system--controller unitary
\begin{align}\label{Usum}
U = \exp\biggl(-i \lambda \AR{\sum_{k=0}^{m-1}} S_k\biggr) = \AR{\bigotimes_{k=0}^{m-1}} \exp(-i\lambda S_k/m),
\end{align}
{which corresponds to a Hamiltonian containing only pairwise interactions between qubits.} If the initial states of the system and the controllers are both separable, then the system and controller qubit pairs can be considered separately and the controllability of the system follows from that of its constituting qubits. The corresponding Kraus operators then satisfy Theorem~1. This holds also if the system starts in an entangled state, since the Kraus operators do not depend on the initial state of the system. For entangled target states, which are carried by the controllers, this is in general not the case. 

\begin{figure}\centering
\includegraphics[width=0.8\linewidth]{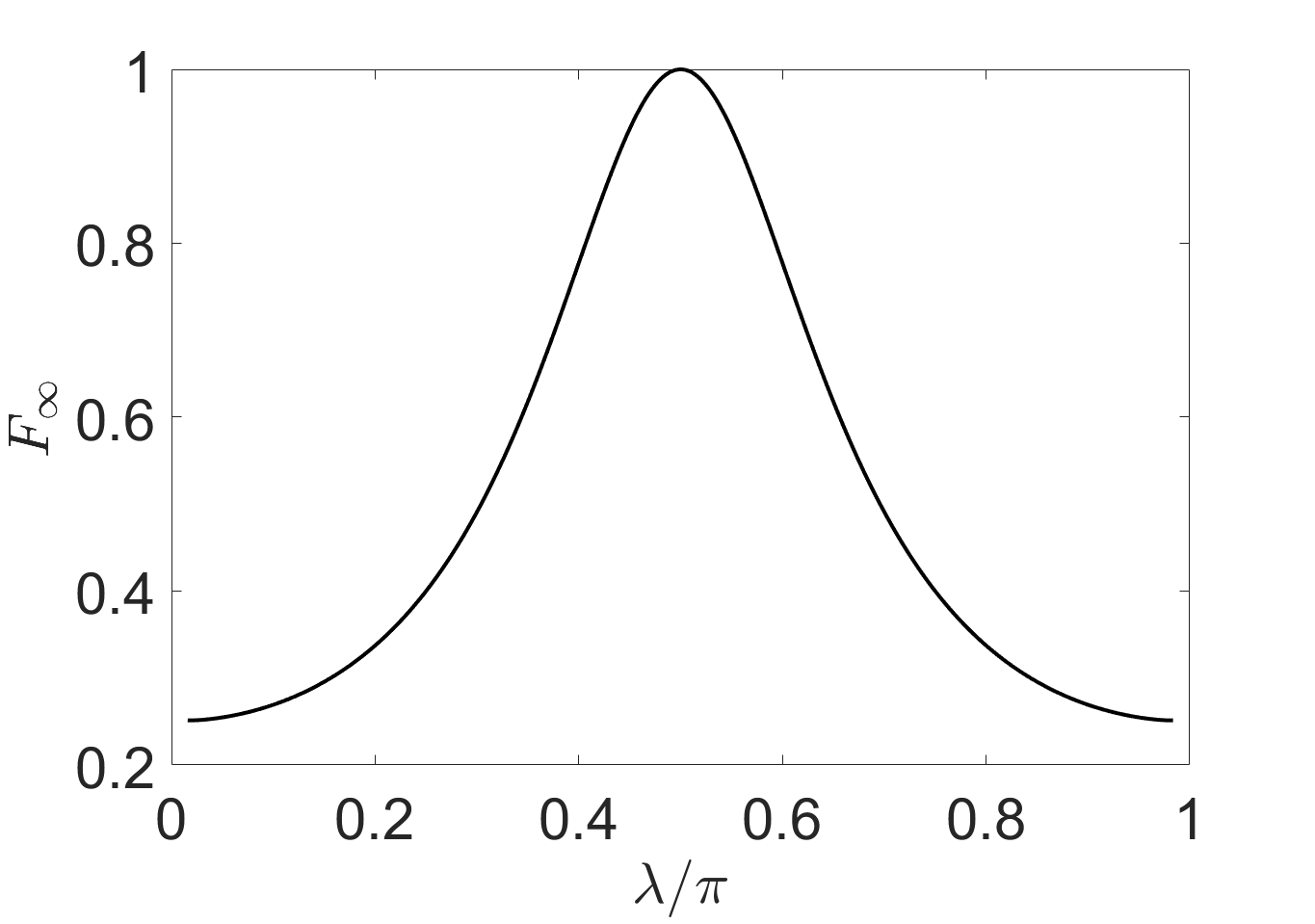}
\caption{\label{fig:fid_epsilon}
{Analysis of the ability of the coupling unitary \eqref{e:U_sum} to reach the target state \AR{$\ket{T} = (\ket{00}+\ket{11})/\sqrt{2}$} exclusively with qubit pair interactions.}
The asymptotic target fidelity \eqref{e:F_inf} is plotted as a function of the parameter $\lambda$ in units of $\pi$ and for a system initial state \AR{$\ket{\psi} = \ket{00}$}. 
}
\end{figure}

To analyse whether, for nonseparable  target states, \eqref{Usum} can still be used to {\em approximately} control a quantum system, we study a system of two qubits, coupled to two-qubit controllers. A straightforward calculation yields
\begin{subequations}
\begin{align}
(S_1 + S_2)^{2n} &= 2^{2n-1} (\mathbbm{I} + S_1 \otimes S_2),\\
(S_1 + S_2)^{2n+1} &= 2^{2n} (S_1 + S_2)
\end{align}
\end{subequations}
for all $n \in \mathbb{N}$, and therefore
\begin{equation}\label{e:U_sum}
U = \cos^2(\lambda)\mathbbm{I} - \sin^2(\lambda)S_1 \otimes S_2 -\tfrac{i}{2} \sin(2\lambda) (S_1+S_2).
\end{equation}
For $\lambda=\pi/2$, this expression simplifies to $U = -S_1 \otimes S_2$, corresponding to a swap that transfers the target state from the controllers to the system.  In order to {probe whether the unitary \eqref{e:U_sum} drives the system at least approximately towards the desired target state also for other values of $\lambda$}, we numerically study the case of an entangled target state. Figure~\ref{fig:fid_epsilon} shows, as a function of $\lambda$, the asymptotic target fidelity 
\begin{equation}\label{e:F_inf}
F_\infty=\lim_{ {n}\to\infty}F(\$^{ {n}}(\rho),T)
\end{equation}
of the channel $\$$ {that corresponds, via Eq.\ (4), to the unitary \eqref{e:U_sum} and controller initial states \AR{$\ket{\psi_0}_c=(\ket{00}_c+\ket{11}_c)/\sqrt{2}$}.} While unit fidelity is reached only at $\lambda= \pi/2$, fairly high fidelities can still be reached in the vicinity of that value, which may permit approximate quantum control by means of purely pair-interacting Hamiltonians, even for strongly entangled target states. 


\section{Control Hamiltonian for \ar{entanglement} creation}
Evaluating the swap operator \eqref{e:swap} for the system basis (25) and the controller basis
\begin{equation}
\AR{\ket{0}_c=\ket{\uparrow\uparrow}_c,\;\ket{1}_c=\ket{\uparrow\downarrow}_c,\;\ket{2}_c=\ket{\downarrow\uparrow}_c,\;\ket{3}_c=\ket{\downarrow\downarrow}_c},
\end{equation}
and expressing the result in terms of Pauli operators, one obtains the Hamiltonian
\begin{align}
H=& \ar{\frac{1}{4}}\left( 
\sigma_2^x\sigma_3^x  +\sigma_2^y\sigma_3^y\sigma_4^x +\sigma_2^z\sigma_3^z\sigma_4^x +\sigma_1^x\sigma_3^x\sigma_4^z \right.\nonumber \\
& + \sigma_1^x\sigma_2^x\sigma_4^z -\sigma_1^x\sigma_2^y\sigma_3^z\sigma_4^y
+ \sigma_1^x\sigma_2^z\sigma_3^y\sigma_4^y -\sigma_1^y\sigma_3^x\sigma_4^y \nonumber \\
& - \sigma_1^y\sigma_2^x\sigma_4^y -\sigma_1^y\sigma_2^y\sigma_3^z\sigma_4^z 
+ \sigma_1^y\sigma_2^z\sigma_3^y\sigma_4^z+\sigma_1^z\sigma_4^x\nonumber \\
& \left.  + \sigma_1^z\sigma_2^x\sigma_3^x\sigma_4^x +\sigma_1^z\sigma_2^y\sigma_3^y
+ \sigma_1^z\sigma_2^z\sigma_3^z + 1 \right)\, .
\end{align}
The subscripts 1 and 2 refer to Pauli operators acting on the first, respectively second, factor of the system Hilbert space $\mathscr{H}_s=\mathbb{C}^2\otimes\mathbb{C}^2$. Analogously, subscripts 3 and 4 refer to operators acting on the first, respectively second, factor of the controller Hilbert space. The corresponding Kraus operators read
\begin{align}
M_1&=\cos(\lambda)\mathbbm{1}-\frac{i}{4}\sin(\lambda)\left(\sigma_1^x \sigma_2^x-\sigma_1^y \sigma_2^y+\sigma_1^z \sigma_2^z+\mathbbm{I}\right),\label{M1}\\
M_2&=-\frac{1}{4}\sin(\lambda) \left(\sigma_1^x \sigma_2^y + \sigma_1^y \sigma_2^x+i\sigma_1^z+i\sigma_2^z\right),\\
M_3&=\frac{1}{4}\sin(\lambda) \left(\sigma_1^y \sigma_2^z+ \sigma_1^z\sigma_2^y - i\sigma_1^x-i\sigma_2^x\right),\\
M_4&=\frac{i}{4}\sin(\lambda) \left(\sigma_1^z \sigma_2^x-\sigma_1^x \sigma_2^z-i\sigma_1^y+i\sigma_2^y\right).\label{M4}
\end{align}


\bibliography{xxbib}